\documentclass[10pt,journal,compsoc]{IEEEtran}

\ifCLASSOPTIONcompsoc
  \usepackage[nocompress]{cite}
\else
  \usepackage{cite}
\fi

\usepackage{amsmath,amssymb,amsfonts}
\usepackage{algorithmic}
\usepackage{graphicx}
\usepackage{textcomp}
\usepackage{xcolor}
\usepackage{color,soul}
 \usepackage{multirow}
 \usepackage{subcaption}
 \usepackage{enumitem}

\usepackage{shortcut}
 
\usepackage{braket}

\usepackage{caption}

\usepackage{url}

\begin{document}
\bstctlcite{IEEEexample:BSTcontrol}
\title{A Systematic Methodology to Compute the \\ Quantum Vulnerability Factors \\ for Quantum Circuits} 
\author{
\IEEEauthorblockN{Daniel Oliveira\IEEEauthorrefmark{2}, Edoardo Giusto\IEEEauthorrefmark{1},
Betis Baheri\IEEEauthorrefmark{3}, Qiang Guan\IEEEauthorrefmark{3}, 
Bartolomeo Montrucchio\IEEEauthorrefmark{1}, Paolo Rech\IEEEauthorrefmark{1}}\\

\IEEEauthorblockA{\IEEEauthorrefmark{2}Department of Informatics, Federal University of Paran\'{a} (UFPR), Curitiba, Brazil\\
dagoliveira@inf.ufpr.br
}\\
\IEEEauthorblockA{\IEEEauthorrefmark{1}DAUIN, Politecnico di Torino, Torino, Italy\\
\{edoardo.giusto, bartolomeo.montrucchio, paolo.rech\}@polito.it
}\\
\IEEEauthorblockA{\IEEEauthorrefmark{3}Department of Computer Science, Kent State University, Kent, USA\\
\{bbaheri, qguan\}@kent.edu 
}\\

}


%

\IEEEtitleabstractindextext{%
\begin{abstract}
Quantum computing is one of the most promising technology advances of the latest years. Once only a conceptual idea to solve physics simulations, quantum computation is today a reality, with numerous machines able to execute quantum algorithms.
One of the hardest challenges in quantum computing is reliability. Qubits are highly sensitive to noise, which can make the output useless. Moreover, lately it has been shown that superconducting qubits are extremely susceptible to external sources of faults, such as ionizing radiation.
When adopted in large scale, radiation-induced errors are expected to become a serious challenge for qubits reliability. 


In this paper, we propose an evaluation of the impact of transient faults in the execution of quantum circuits. 
Inspired by the Architectural and Program Vulnerability Factors, widely adopted to characterize the reliability of classical computing architectures and algorithms, we propose the Quantum Vulnerability Factor (QVF) as a metric to measure the impact that the corruption of a qubit has on the circuit output probability distribution. First, we model faults based on the latest studies on real machines and recently performed radiation experiments. Then, we design a quantum fault injector, built over Qiskit, and characterize the propagation of faults in quantum circuits. We report the finding of more than 15,000,000 fault injections, evaluating the reliability of three quantum circuits and identifying the faults and qubits that are more likely than others to impact the output.
With our results, we give guidelines on how to map the qubits in the real quantum computer to reduce the output error and to reduce the probability of having a radiation-induced corruption to modify the output. 
Finally, we compare the simulation results with experiments on physical quantum computers.
\end{abstract}
}

\maketitle

\IEEEraisesectionheading{\section{Introduction}\label{sec:introduction}}

\label{sec_intro}

Quantum computing is quickly moving from being a conceptual solution to physics problems to an extremely efficient and promising computing architecture for critical applications, such as big data~\cite{1996Grover}, 
machine learning~\cite{lloyd2013quantum}, 
chemistry~\cite{Peruzzo2014}, and drug development~\cite{quantum-drug}, just to name a few. 

The turning point that made the intriguing theory of quantum computing a promising computing paradigm was the achievement of sufficiently fault tolerant qubits to allow the computation of small, yet crucial, circuits~\cite{Bravyi2018, Chamberland_2020}. 
The reliability challenge of qubits is intrinsic in the unpredictability of quantum mechanics (the state of qubits can be randomly changed) and the sensitivity of qubits to external perturbations. 
As technology improved, major industries developed their prototypes of quantum machines and today offer researchers access to several quantum computers, such as IBM, D-Wave, Rigetti, Pasqal, and quantum circuit  simulators~\cite{Qiskit, AngliQSimulator}. 
The billions of dollars investments of industries, research centers, and government agencies in quantum computing are encouraging the development of large-scale quantum computers as well as the training of quantum programmers and designers.

Recently, pioneer works have demonstrated that it is necessary to harden superconducting qubits also from external radiation~\cite{radiation2011, Cardani2021, Martinis2021, Chen2021} as the interaction of ionizing particles significantly reduces the fault tolerance of qubits~\cite{LossMechanisms2018, nature_rad, muons2021}. Trapped-ion qubits are found to be robust to low-dose low-energy radiation~\cite{trappedionlowdose}, but no data is available for heavier particles, yet. 
Already being one of the challenges for today's classical computing systems, then, ionizing radiation is expected to be a major issue also for future quantum (super) computers~\cite{Cardani2021, Chen2021}.
Actually, as qubits have a higher sensitivity to external perturbation than CMOS transistors, quantum computers might be even more susceptible to ionizing radiation than classical computers. Recent studies showed that qubits can be affected by light particles, such as muons~\cite{muons2021} or even infrared light~\cite{Barends2011}, that do not have sufficient energy to significantly impact CMOS behavior. 

Despite the fact that quantum computers are not yet fully available in a large scale of qubits, understanding and mitigating the radiation-induced faults is not premature. 
Recent discoveries have urged substantial reductions in operational error rates and further research into the mitigation of error mechanisms such as high-energy particles~\cite{Cardani2021, Chen2021}. 
We cannot risk vanishing the efforts in producing stable quantum computers for realizing then, once adopted in the field, that the radiation-induced error rate is too high and still have not yet fully understood fault propagation.  

In this paper, we aim at anticipating the need for a formal metric and providing researchers and developers with practical tools to evaluate the reliability of quantum circuits. We investigate the effect on the quantum output of faults affecting each qubit of the circuit.
Taking inspiration from the Architectural and Program Vulnerability Factors (AVF~\cite{Mukherjee2003} and PVF~\cite{PVF}), which are the two most widely used metrics to measure the reliability of a device or code, we define the \emph{Quantum Vulnerability Factor} (QVF) as an indicator of the vulnerability of the circuit to faults. A circuit with a low value of QVF is less vulnerable to faults than a circuit with a high QVF. %
Similarly to AVF and PVF, and for the same reasons, we do not investigate the probability for the fault to occur, but we rather assume that the fault occurred and track its propagation.
The fault occurrence probability and fault generation mechanisms study is extremely interesting as it allows to estimate the fault rate of a system.
Nevertheless, as radiation-induced faults are stochastic~\cite{Baumann2005}, the fault probability does not help the circuit/code designer or the architect to improve the reliability of the system.
On the contrary, identifying the resources that, once corrupted, are more likely to impact the output correctness allows to quickly understand the vulnerabilities of the system and eventually take proper countermeasures. 
This applies to classical hardware resources (AVF), code portions (PVF), and, as we show in this paper, also to qubits and quantum circuits (QVF).

\begin{figure*}[t]%
   	\centering
    	\includegraphics[width=0.8\textwidth]{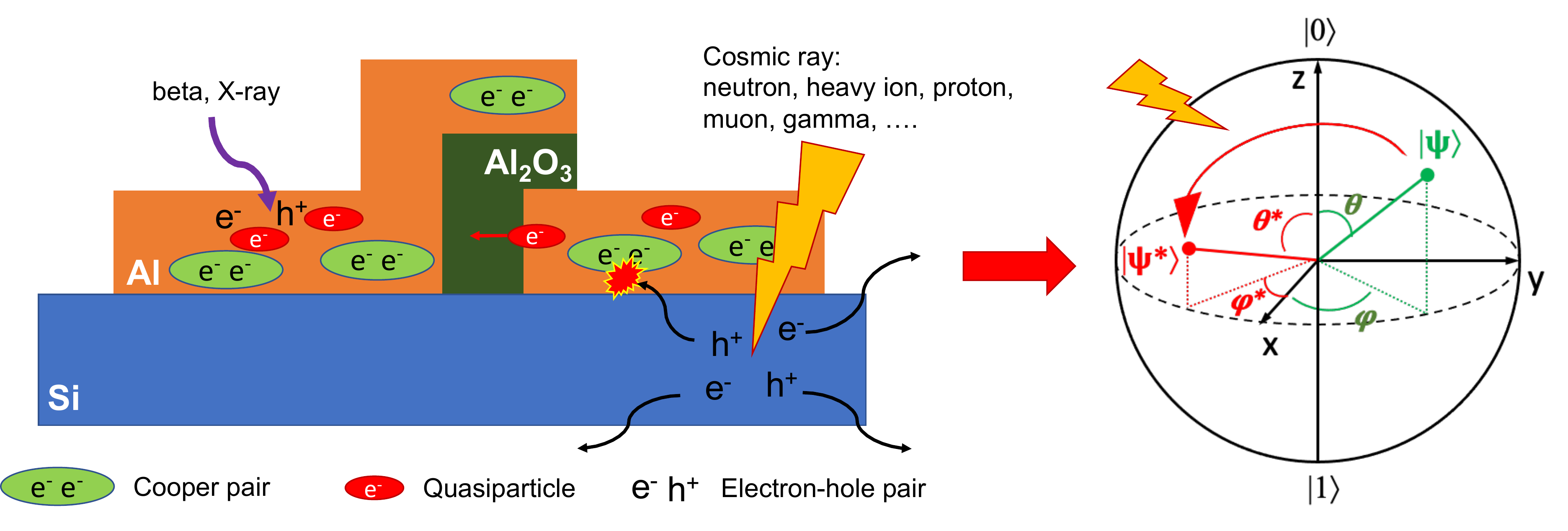}
        \caption{Simplified effect of the impact of ionizing radiation with superconducting qubit material (on the left, adapted from~\cite{nature_rad}) and its consequences on the qubit state (on the right, adapted from~\cite{muons2021}). The charge deposited by radiation in Silicon or Aluminum generates electron-hole pairs that induce a non-equilibrium and breaks the Cooper pair, producing quasiparticles. The excess of quasiparticles excites the qubit. The excitement is logically translated into phase(s) shift(s) and a state change from the expected $| \Psi \rangle$ to the corrupted $| \Psi ^\ast \rangle$.}
        \label{fig_rad_effect}
\end{figure*}

The main contributions of our effort are:

\begin{itemize}[noitemsep,topsep=0pt]
\item Based on the latest studies, we define how to model radiation-induced faults in qubits. Qubits do not have a binary value, thus, simply flipping a bit is not sufficient to model a fault.
The status of a qubit is described by polar coordinates on the Bloch sphere (see Figure~\ref{fig_rad_effect}), and any modification to its value can be represented as a shift in one or both angles of the representation.
We inject different phase shifts of different magnitudes to identify which phase shift is more critical and to highlight any possible correlation between the magnitude of the shift and the impact on the output.
\item We design a quantum fault injector built on top of Qiskit that can also run on real quantum machines.
We inject a fault by introducing a $U$ gate in the target qubit.
For each qubit, we inject $312$ state changes in several positions of the circuit. The fault injector, the circuits, and the whole fault injection campaign are publicly available~\cite{REPO}. 
\item To study the fault impact in the output probabilities we introduce the QVF metric, based on the Michelson Contrast, that quantifies how much the fault in a qubit reduces the confidence of the result.
\item We provide a detailed QVF evaluation, identifying the quantum circuits and the qubits in each circuit that are more likely to be corrupted and quantifying the amplitude of the fault that is sufficient to corrupt the output.
\end{itemize}


The rest of the paper is organized as follows.
To define, show a practical application, and prove the importance of QVF in understanding the reliability of quantum circuits, we give, in Section~\ref{sec_background}, some background information about quantum computing and radiation effects in qubits.
Then, in Section~\ref{sec_qvf}, we formalize the Quantum Vulnerability Factor metric and, in Section~\ref{sec_framework}, we describe the fault model we inject and the fault injection framework we used to evaluate the QVF of qubits and quantum circuits.
The obtained results are presented in Section~\ref{sec_results}, in Section~\ref{sec_discussion} we discuss the impact of our findings, and Section~\ref{sec_conclusion} concludes the paper.





\section{Background and Related Work}
\label{sec_background}

In this Section, we give a background about quantum computing and the impact of ionizing radiation on qubits, which is essential to understand and appreciate the contribution of our paper.
As there is not yet sufficient data on radiation effects on trapped-ion qubits (besides the low-dose test documented in~\cite{trappedionlowdose}), we will focus the discussion on superconducting materials. Nonetheless, the concepts we introduce, the fault injection framework we design, and the impact of the results we present are independent on the qubit technology, once the fault model is defined. The details provided in this Section serve as a solid background to justify the chosen fault model.

\subsection{State of the art in quantum computing systems}



Even though the physical implementation of quantum computing technology covers more than 20 different approaches from Quantum dot computers to Nuclear magnetic resonance quantum computers, the most popular and successful attempts in this domain are Trapped Ion and Superconducting quantum computers, the latter being capable of running the majority of quantum algorithms.


As the biggest quantum computing service provider, the IBM quantum system includes 21 quantum computers available, with qubit capacity ranging from 5 to 65, with public access through API services provided by IBM. Among the 21 available quantum computers, three types of quantum processors are implemented: Canary, Falcon, and Hummingbird. Each quantum processor model uses different qubit topologies, which may greatly impact the machine quantum noise and quantum errors on computation results~\cite{qucloudHPCA}.

Quantum programs are expressed as quantum circuits, in which a set of quantum gates are sequentially applied to the initial qubits, and produce a probabilistic output across all combinations of the classical bits. These quantum circuits will be mapped to qubits from Noisy Intermediate-Scale Quantum (NISQ) machines based on the topological connectivity of the qubits and the availability of the supported gate types.
Programming on different quantum platforms and quantum simulators has been constrained by the quantum compilers supported by different systems. Qiskit~\cite{Qiskit} and QASM~\cite{cross2021openqasm} are generally supported by IBM quantum machines. Both Qiskit and QASM are imperative programming languages and designed machine-independent. In our experiments, we mainly focus on studying the IBM quantum machines and running benchmark quantum programs in Qiskit.

\subsection{Quantum noise}
\label{sec_noise}

In spite of the fact that the qubit capacity of quantum computers is ever-increasing, the performance of IBM quantum computers is bounded by the quantum noise, which is greatly reducing the accuracy and performance of the IBM quantum computers~\cite{Harper2020}.
Quantum noise can be categorized into operation errors and retention (coherence) errors~\cite{2017APSMARR51007G}. A qubit can retain data (position) only for a limited time (coherence time). Retention errors are categorized into two types, T1 and T2 errors~\cite{Hu01Decoherence}. 
A qubit in a high energy state $\ket{1}$ naturally decays to lower energy state $\ket{0}$, the time associated with this decay is called spin-lattice coherence time (T1). The spin-spin relaxation process (T2) indicates the time for a qubit to get affected by the external environment or by the interaction with other qubits. 
Depending on the material used in NISQ machines, individual qubits have a time range for T1 and T2 that has improved  in the last decade from 1 nano-second to 100 micro-seconds~\cite{Kjaergaard_2020}.

Hardware-based noise reduction is still under investigation by IBM-Q, which is working to improve the environment and qubit design by updating the quantum processor and even using new isolation techniques. In the IBM-Q system One, IBM tried to isolate the interaction with noise and qubits by separating the quantum computer from the surrounding environment.
However, the hardware approach is only one of the dimensions to consider, depending on the circuit depth and noise on gates and measurements, the level of quantum noise affecting the results can change in a certain range.

Quantum Error Correction (QEC) has been fundamental to reach Fault Tolerant Quantum Computers (FTQC). QEC is designed to protect a qubit from noise, at the extremely high cost of requiring from 5x to 9x larger circuits. As particles impacts, unlike noise, are stochastic and unpredictable, unfortunately, QEC is inefficient in handling radiation-induced and correlated faults~\cite{nature_rad, muons2021, LossMechanisms2018}. To reach FTQC we need to design better QEC, and the first step is the understanding of faults impact and propagation, which is the main scope of this paper.


\subsection{Radiation-Induced Faults in qubits}
\label{sub_radiation}

Preliminary and inspiring works show that ionizing radiation induces faults in superconducting qubits~\cite{LossMechanisms2018, nature_rad, muons2021, Barends2011} and, once employed in large scale, radiation fault tolerance is expected to be the next big challenge for quantum (super-) computers~\cite{radiation2011, Cardani2021, Martinis2021, Chen2021}.

As shown in Figure~\ref{fig_rad_effect}, adapted from~\cite{nature_rad} and~\cite{muons2021}, the impact of ionizing radiation on the qubit superconducting materials increases the amount of hole-electron pairs in the Aluminum thin-film and Silicon substrate. Heavy particles are more likely to interact with Silicon ($Al$ is transparent to neutrons), other sources of radiation ($\beta$, X-rays) with Aluminum.
While there is still no quantitative measurement of the radiation-induced fault rate in quantum circuits,
it is known that the additional charge deposited by the impinging particle induces a non-equilibrium that leads to Cooper pairs break and, thus, quasiparticles generation~\cite{nature_rad, muons2021}. The resulting excitement modifies the state of the qubit, possibly changing its state (i.e., it induces $\phi$ and/or $\theta$ phase shift) as shown with simulations~\cite{muons2021} and experimentally validated~\cite{Cardani2021}. While for CMOS a fault is generated only if the charge is higher than the critical charge~\cite{Baumann2005}, on a qubit any excitement modifies the state, inducing a phase shift with a magnitude that depends on the deposited charge~\cite{Catelani2011}. 
It has also been shown that if, and only if, the deposited charge is sufficiently high the qubit collapses~\cite{nature_rad}. 

Terrestrial neutrons and heavy ions, which are today the most critical source of faults for Silicon-based classical computing devices~\cite{Baumann2005}, generate a large amount of electrons-holes pairs in the Silicon substrate. The energy spectrum of neutrons ranges from meV to GeV, the lower energy ones being exponentially more common.
Unfortunately, the frailty of qubits makes it not unlikely for low energy neutrons and lighter particles, such as muons (almost harmless for CMOS technology~\cite{muons2011}), to induce a sufficient perturbation in the qubit to generate a fault~\cite{muons2021}. 

An interesting difference between particles is that $\gamma$-rays, $\beta$, X-rays have a constant and accumulative effect while the neutrons, heavy ions, and muons impact is stochastic and transient. In other words, $\gamma$-rays, $\beta$, and X-rays exposure constantly deposits a little amount of charge, until inducing a fault or qubit collapsing~\cite{nature_rad} (or a permanent CMOS transistor malfunction~\cite{Oldham2003}). Neutrons and heavy ions strikes are random (the neutrons flux at sea level is $\sim13 n/(cm^2\times{h})$~\cite{Jedec2006}) and the impact with the device material produces a transient charge that can lead to a fault.

Unfortunately, while a thin shielding could be sufficient to drastically reduce the number of X-rays reaching the qubit, as mentioned in~\cite{nature_rad}, the shielding for neutrons and heavy ions is impractical (meters of concrete or lead) and for muons is basically impossible as the qubit should be placed in deep underground caves~\cite{muons2021}. As a result, as known for traditional computing devices, it is impossible to shield qubits from transient faults. We must find effective and efficient solutions to deal with the unavoidable transient faults in qubits.

\subsection{Contribution}
\label{sec_contrib}
Unlike previous work on quantum fault tolerance, we focus on the effects of faults in quantum computation rather than investigating the fault physical mechanisms. As in the AVF or PVF measurement, we assume that the fault occurred independently of the cause, and understand its effect on the quantum circuit output.
While recently a preliminary fault injector to track noise propagation was presented~\cite{Resch2021}, there is still no fault injector to track transient fault propagation in quantum circuits. In this paper, we introduce a novel fault injector integrated with Qiskit and that can be used also in real quantum machines. Additionally, we formalize a new metric, the Quantum Vulnerability Factor, based on Michelson Contrast, to ease the analysis of quantum circuit reliability.

We aim at practically estimating the reliability of a circuit and at identifying the qubits in a circuit that, once affected, are more likely to induce a negative impact on the circuit correctness. Such information is highly valuable as it allows to map the circuit qubits to physical qubits in the most reliable way and to predict the effects of faults in the quantum computation. 





\section{Quantum Vulnerability Factor}
\label{sec_qvf}





In this Section we present the Quantum Vulnerability Factor (QVF) metric, to better understand the reliability and fault propagation in quantum circuits.
As we will discuss, while there are already some metrics available to quantify the quality of quantum circuit outputs, none of these metrics provide sufficient information on the fault propagation effect.


\subsection{Definition}



The output of a quantum circuit is a set of states, each with a different probability. The state(s) with the higher probability is(are) considered the output.
Most works use the Probability of Successful Trial~(PST) metric when evaluating the reliability and correctness of a quantum circuit output~\cite{10.1145/3297858.3304007, tannu2019mitigating, liu2021qucloud, das2019case}. 
PST considers only the probability of the correct state and is defined by Equation~\ref{eq_pst}. 

\begin{equation}
    PST = \frac{Number\ of\ successful\ trials}{Total\ number\ of\ trials}
        \label{eq_pst}
\end{equation}

\begin{figure}[t]%
   	\centering
    	\includegraphics[width=0.98\columnwidth]{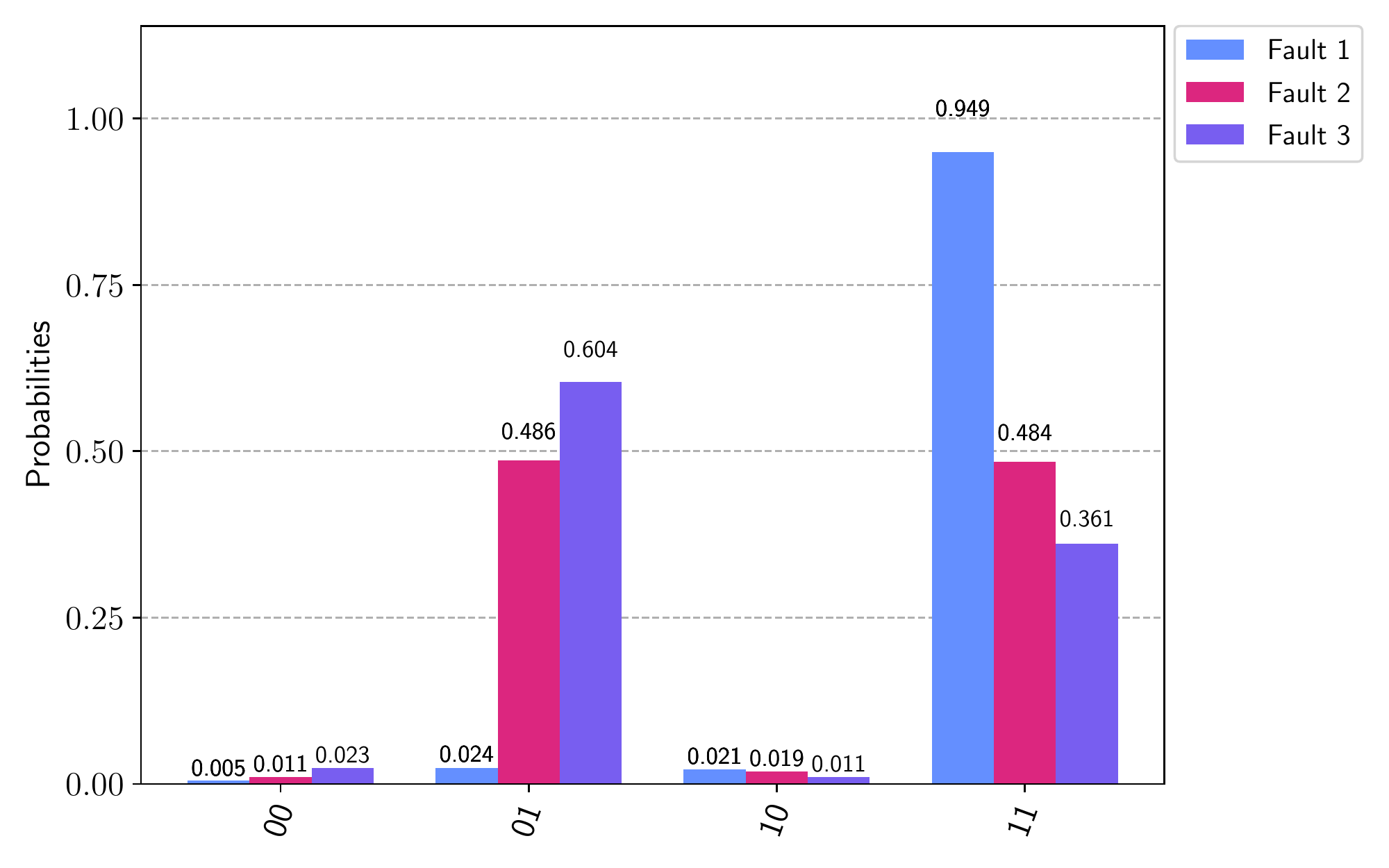}
        \caption{The probability distribution for three fault injections in a 2-qubit Grover circuit. The correct state is $11$, and QVF resulting from the injection of fault 1, 2, and 3 is $0.03$, $0.50$, and $0.63$ respectively.}
        \label{fig_qvf_example}
\end{figure}

The PST metric on its own cannot quantify nor qualify if a circuit is reliable and cannot give overall information about the state probability distribution we are investigating. To use PST one needs to  specify a threshold to define when a circuit can be considered reliable. For instance, one could argue that circuits with $PST > 0.68$ are sufficiently good, and a circuit with a $PST = 0.5$ fails the reliability assessment. Unfortunately, defining a threshold, such as 1-sigma, masks the details of the fault effects on the output state distribution and may deem reliable circuits as unreliable. As we show later in this Section, using the example in Figure~\ref{fig_qvf_example}, values close to the threshold are identified as correct but can be highly susceptible to external perturbations, such as radiation.
While increasing the PST may indeed improve the reliability, in fact, this increase can also affect the overall probability distribution and thus increase the probability of a single incorrect state. 







To understand the reliability of a circuit, and consequently the impact of a fault on the output correctness (i.e., on the output probability distribution), we extend the PST metric as follows. First, we compute the Michelson Contrast~\cite{kukkonen1993michelson}.
We use the Michelson Contrast since it naturally measures how distinguishable one object is from others using color, luminance, or, as in our specific case, the probability of each output state.
In other words, the Michelson Contrast defines how confidently (i.e., how distinguished) one can select the correct state among all states in the output. 

\begin{equation}
    Contrast = \frac{P(A) - P(B)}{P(A) + P(B)}
    \label{eq_contrast}
\end{equation}

Equation~\ref{eq_contrast} shows the contrast computation, where $P(A)$ is the probability of the correct state (i.e., the expected state in a fault-free execution), and $P(B)$ is the highest probability among any of the incorrect states (i.e., the most probable incorrect state).
Please note that PST is precisely $P(A)$ in Equation~\ref{eq_contrast} of our contrast metric~\cite{tannu2019mitigating}, but we include $P(B)$ to consider also the overall probability distribution.

Since it is possible for a circuit to produce $P(A) < P(B)$ (e.g., due to technology, noise, or external factors such as radiation-induced faults), the contrast range is $[-1,1]$.
It is worth noting that the contrast metric is not limited to circuits with a single correct state, the extension for multiple-state circuits can be easily performed by aggregating the probabilities of all correct states into $P(A)$.


To shift the range to $[0,1]$ and to have lower values indicating a more reliable configuration (as for AVF and PVF), the QVF is calculated as shown in Equation~\ref{eq_qvf}:

\begin{equation}
    QVF= 1 - (Contrast + 1) /2
        \label{eq_qvf}
\end{equation}

By this definition, QVF values close to zero indicate a clear contrast between the correct state and the incorrect ones, with the correct state presenting the highest probability.
In other words, the probability to have the expected output state is very high compared to the other states.
QVF values around $0.5$ present the correct state and at least one incorrect state with similar probabilities, which makes the identification of correct states dubious. Finally, values close to one represent the worst case in a probability distribution where the correct states are not even as high as the incorrect ones.

To better illustrate the QVF calculation and meaning, Figure~\ref{fig_qvf_example} shows the probability distribution for a 2-qubits Grover circuit in which we perform 3 fault injections. The correct state is $11$ and the most probable incorrect state is $01$ for all three faults. Then, $P(A) = P(11)$ and $P(B) = P(01)$ for all three fault injection. The $QVF$ results for each circuit are as follows:
\begin{itemize}
    \item Fault 1: $P(A) = 0.949$, $P(B) = 0.024$, $QVF = 0.03$
    \item Fault 2: $P(A) = 0.484$, $P(B) = 0.486$, $QVF = 0.50$
    \item Fault 3: $P(A) = 0.361$, $P(B) = 0.604$, $QVF = 0.63$
\end{itemize}

 The circuit has a very low QVF for the first fault ($0.03$), as evident from Figure~\ref{fig_qvf_example}, indicating that the correct state $11$, with a probability of about 95\%, can be reliably selected. The second fault has a QVF of about $0.50$, which indicates that it is not possible to reliably select the correct state ($11$ and $01$ states are almost equally probable). Finally, the third fault has a high QVF of $0.63$, indicating that the incorrect state $01$ is more likely to be selected as the output, which can lead to errors. 
 As a result, we can state that fault 1 (low QVF) is not very critical, while fault 2 and mostly fault 3 should be strictly avoided as they drastically change the output state distribution. It is worth noting that PST would have identified the outcome of fault 1 as correct and both faults 2 and 3 as wrong. However, no information about the impact on the output state distribution would have been given.
 
\begin{figure}[t]%
   	\centering
    	\includegraphics[width=0.98\columnwidth]{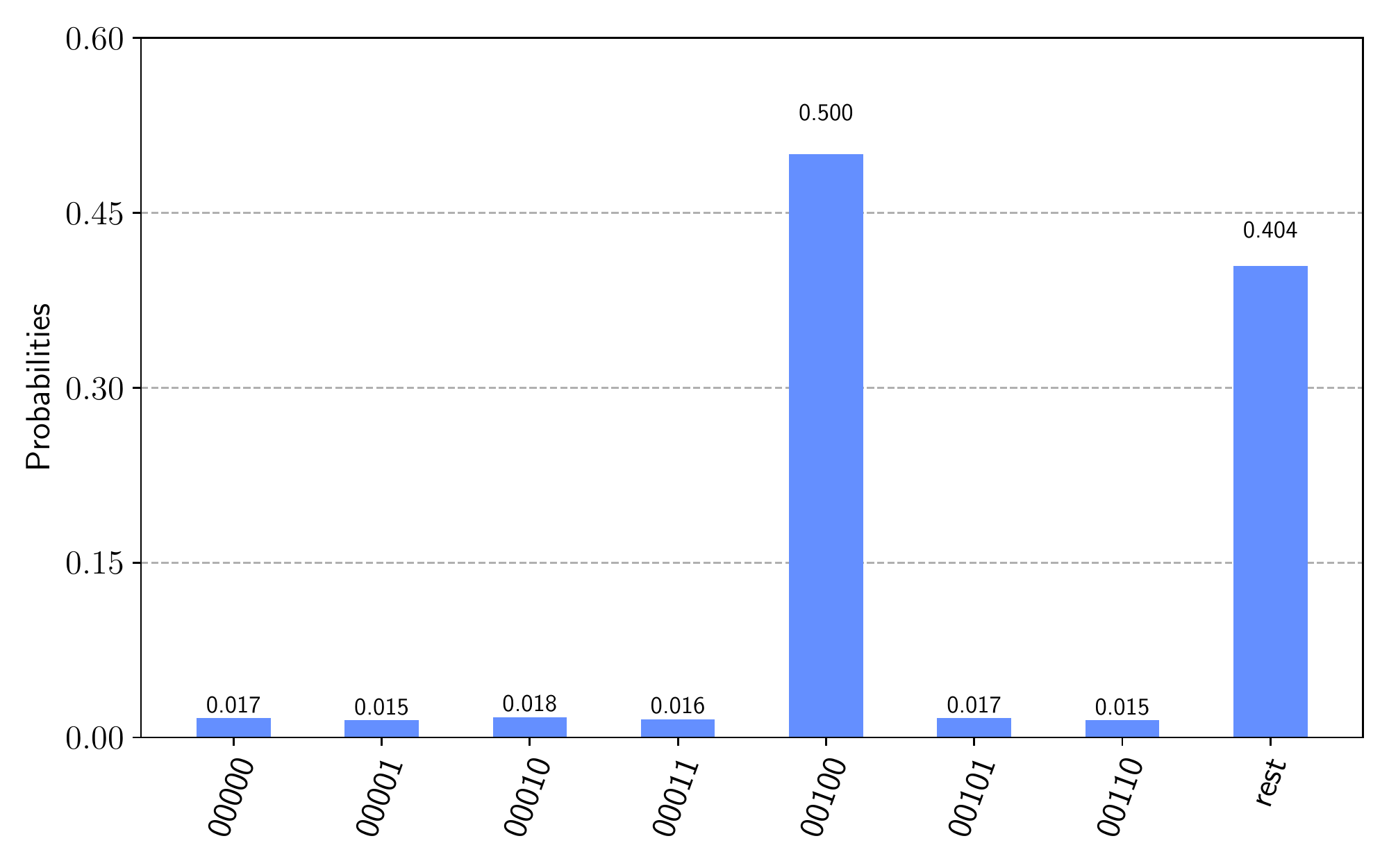}
        \caption{The probability distribution for a fault in a 5-qubit circuit (the rest is the sum of the probabilities of the remaining states). The correct state is $00100$, resulting in $PST = 0.5$ (probably considered wrong) and $QVF = 0.03$ (fault has a minor effect in the output).}
        \label{fig_pst_vs_qvf}
\end{figure}

To further illustrate the weakness of PST with an arbitrary threshold in the evaluation we are proposing, let us consider a fault in a 5-qubit circuit with the probability distribution shown in Figure~\ref{fig_pst_vs_qvf}. 
The correct state of this circuit is $00100$ and the probability for this state is about 50\%, while the probability for any incorrect state is at most 1.76\% (i.e., all of the 31 remaining states have a similar probability).
Thus, the output of this faulty circuit results in a PST of $0.5$, rejecting that circuit as unreliable (depending on the threshold defined).
However, the QVF is $0.03$ indicating that the correct state can be confidently identified, even if one defines a strict threshold.
This is evident in Figure~\ref{fig_pst_vs_qvf}, since the state $00100$ is the only state with non-negligible probability, and the remaining states can be easily identified as noise.



To measure the QVF as well as the impact of faults, we designed a fault injector for the quantum circuit described in Section~\ref{sec_framework}.
We inject faults in each qubit and track the effect of each injection in the output.
When comparing the QVF of a fault-free circuit to a faulty one, we can observe how much a circuit or its individual qubits are sensitive to faults.

\section{Fault Injection Framework}
\label{sec_framework}

\label{sub_fi}
\begin{figure*}[ht!]%
   	\centering
    	\includegraphics[width=0.98\textwidth]{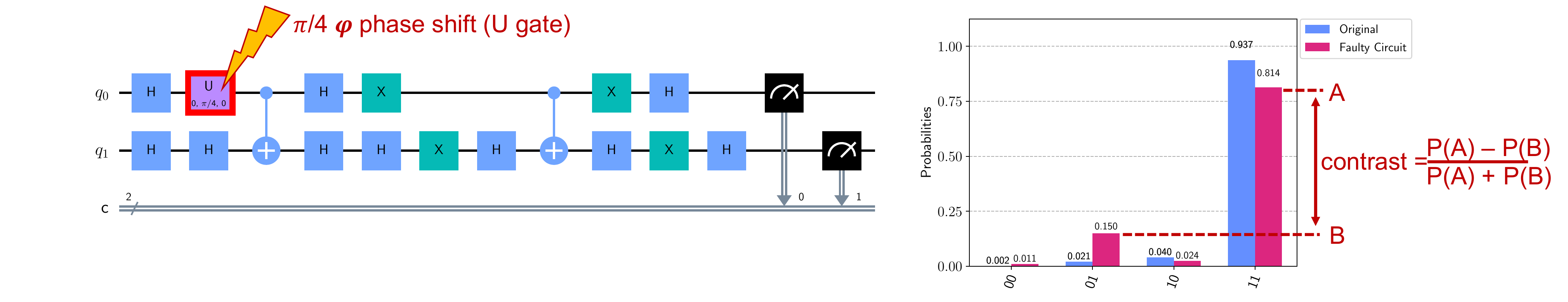}
        \caption{Example of fault injection in the Grover circuit (left) and QVF calculation (right). A $\theta$ shift of $\frac{\pi}{4}$ is injected in $q_0$ after the first H-gate. The fault modifies the output probabilities distribution, shown to the right, from the blue one to the red one. As shown in Equation~\ref{eq_qvf}, QVF is calculated using the Michelson Contrast, where A is the probability of the expected (fault-free) output (11 in this case) and B is the highest probability among the wrong outputs (01).}
        \label{fig_fault_example}
\end{figure*}

In this Section, we first describe how we model the fault from the available knowledge on qubit sensitivity to radiation.
Then, we describe the fault injection framework we developed.

\subsection{Fault Model}
\label{sub_fault_model}

We model the faults to inject based on the latest discoveries on qubits radiation vulnerability, that have shown that even a small deposition of charge from an impinging particle reduces the coherence time and can lead to unexpected modifications of the qubit state~\cite{muons2021, Catelani2011, Cardani2021, nature_rad}.

In a classical computer, a fault is generated when the impinging particle deposits enough charge to change the binary state of a transistor (i.e., the deposited charge is higher than the critical charge~\cite{Baumann2005}). The qubit state is associated with the $\phi$ and $\theta$ coordinates that represent its amplitude and phase. In a qubit, then, the excitement that derives from the impact of the impinging particle is logically reflected as a state change and can be represented with $\phi$ and/or $\theta$ phase(s) shift(s), 
independently of the source (see Figure~\ref{fig_rad_effect}). 

A $\theta$ phase shift changes the 0-1 probability in the qubit, while a $\phi$ phase shift changes its orientation.
Both shifts can impact the correctness of the execution, and it is part of our contribution to identify which phase shift in which qubit is more critical for a quantum circuit.
Albeit not yet experimentally quantified, it has been mathematically shown that a higher charge deposition induces a bigger phase shift~\cite{Catelani2011}.
As the charge deposition of the radiation-induced impact depends on the energy of the impinging particle (which goes from meV to GeV~\cite{Baumann2005}) and on the distance between the impact location and the qubit, the resulting phase shift magnitude can be largely variable. Thus, we inject shifts of different magnitudes as well as combinations of shifts ($\theta + \phi$) to correlate the circuit behavior to a wide spectrum of effects.
As discussed in Section~\ref{sub_fi}, the amplitude and direction of the injected phase shift is a parameter in our fault injection framework.

A major event or a cumulative charge deposition, as shown in~\cite{nature_rad} for X-rays, can also lead the qubit to collapse. In such an event, the qubit ceases to operate, there would be no reason to measure its QVF 
and there would be no other practical solution than re-executing the whole circuit. As the vast majority of neutrons have low energy and the cumulative effect of X-rays can be easily shielded~\cite{nature_rad}, radiation-induced qubit collapses are expected to be less likely than phase shifts.


\subsection{Noise}
\label{sub_noise}

The quantum circuit execution is known to be noisy, even in the absence of faults.
Faults will happen on top of the already noisy execution. 
Based on the information provided e.g., by IBM,
it is possible to have a clear and detailed characterization of the noise in each physical qubit~\cite{9283531}.


To have a full understanding of the effects of faults in the quantum circuit execution, our fault injector can be executed with three different scenarios.
(1) Simulation without external noise, ideal but unrealistic environment.
(2) Simulation of a physical machine, using the IBM-Q noise model to create a realistic environment based on actual quantum computers.
(3) Injection over the circuit execution on a physical IBM-Q machine.
In this paper, we only present data obtained with scenario (2) and (3) since scenario (1) cannot be achieved in the real world. 
Considering the noise, executed or modeled after a physical machine, helps us to understand how each qubit and fault severity can change the state probability distribution. It is worth noting that it is possible for a fault to attenuate the noise effect by changing $\phi$ or $\theta$ in opposite directions, improving the circuit reliability.



\subsection{Fault Injector}
\label{sec_faultinjector}



Our fault injector is built on the open-source and well-documented Qiskit framework~\cite{Qiskit}. The fault injector operates over a Qiskit's \textit{QuantumCircuit} object to generate new circuits with fault(s) injected. 
The new faulty circuits can be transpiled and executed just as one would execute a regular \textit{QuantumCircuit}. Thus, we can execute the faulty circuits in physical IBM-Q machines as well as simulators, or even export them as QASM files to load and execute the circuits on different systems.

To model the injected fault, we used Qiskit's most generic gate, the \textit{U} gate, to effectively simulate every possible phase shift.
The \textit{U} gate can be described as: 
\begin{equation}
U(\theta, \phi, \lambda)=\left[\begin{array}{cc}\cos \left(\frac{\theta}{2}\right) & -e^{i \lambda} \sin \left(\frac{\theta}{2}\right) \\ e^{i \phi} \sin \left(\frac{\theta}{2}\right) & e^{i(\phi+\lambda)} \cos \left(\frac{\theta}{2}\right)\end{array}\right]
\end{equation}
which receives three parameters: 
\begin{itemize}
    \item $\phi$ is the angle defined in the $XY$ plane of the Bloch sphere (a rotation angle on the $Z$ axis);
    \item $\theta$ is the angle defined in the plane that includes the $Z$ axis and the vector representing the generic quantum state $\ket{\psi}$; 
    \item $\lambda$ is also a rotation on the $Z$ axis.
\end{itemize}

The parameters for the $U$ gate have been thus selected:
\begin{itemize}
    \item $\phi$ = $[0, 2\pi[$ each $15{^\circ}$;
    \item $\theta$ = $[0, \pi]$ each $15^{\circ}$;
    \item $\lambda$ = 0
\end{itemize}
This angles combination results in $312$ possible configurations of the $U$ gate (injections) for each position in the quantum circuit.

\begin{figure*}[!ht]%
    \begin{subfigure}{.33\textwidth}
   	    \centering
        \includegraphics[width=\textwidth]{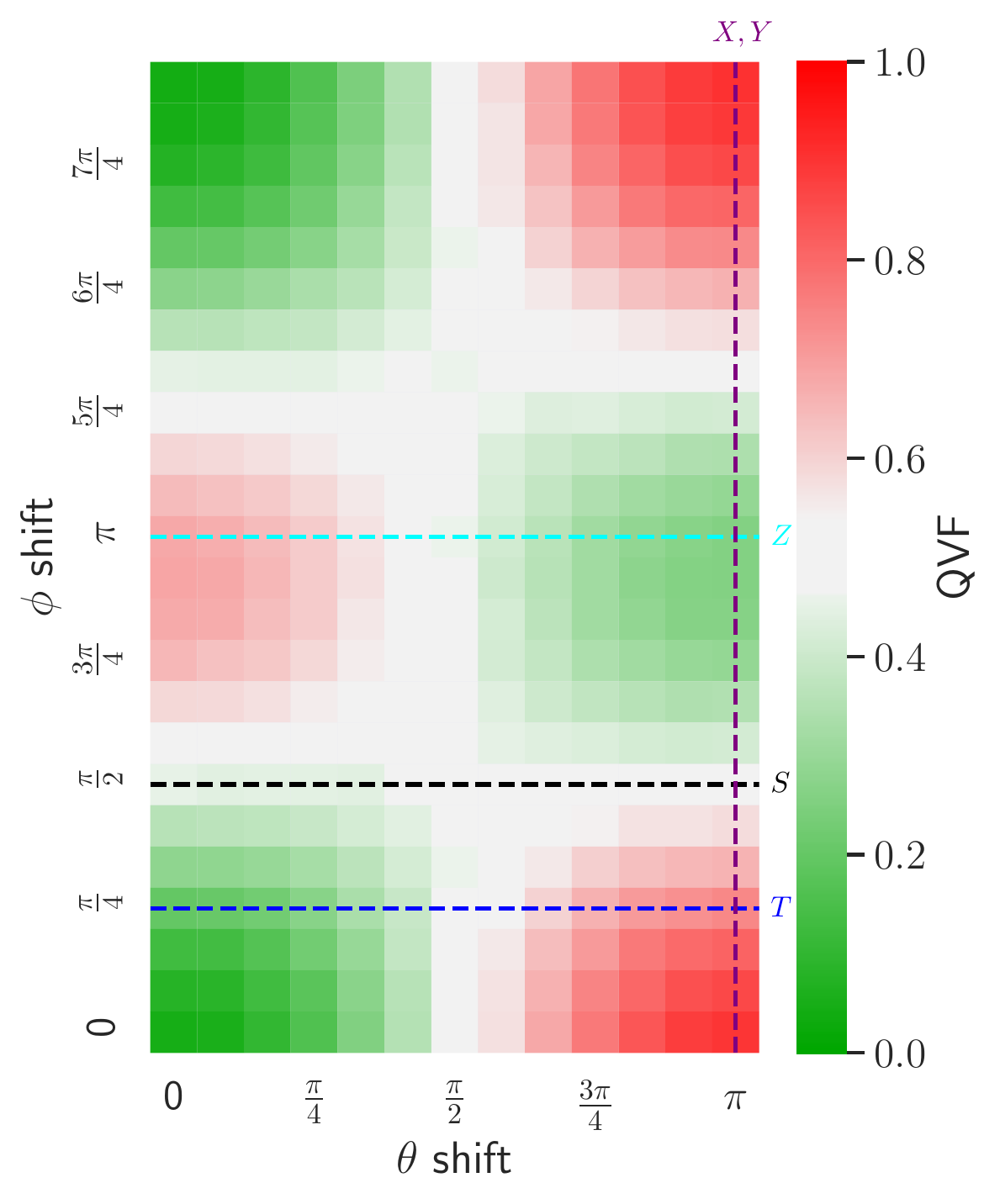}
        \caption{Bernstein-Vazirani.}
        \label{fig_hm_bv}
    \end{subfigure}%
    \hfill
    \begin{subfigure}{.33\textwidth}
        \centering
        \includegraphics[width=\textwidth]{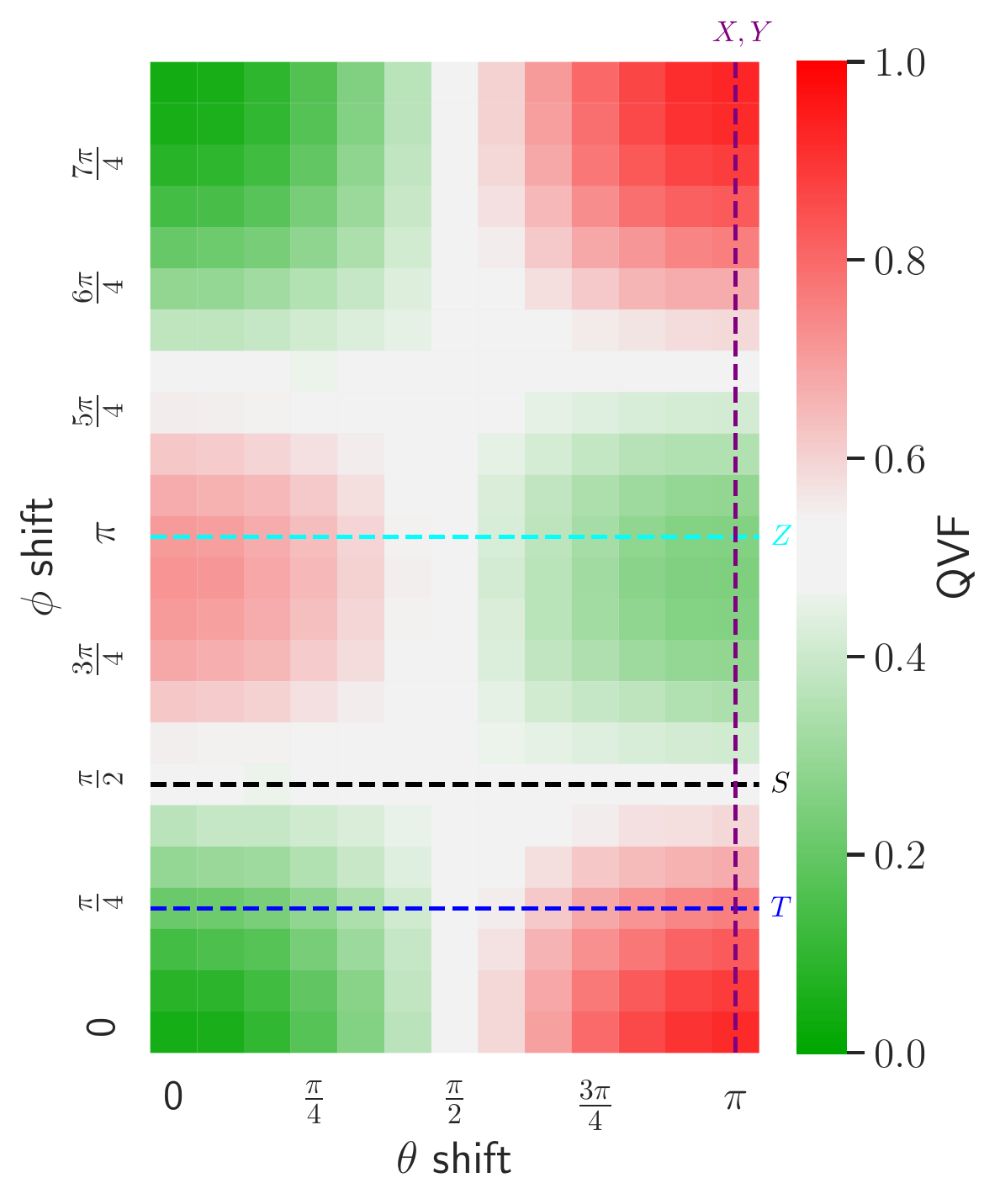}
        \caption{Deutsch-Jozsa.}
        \label{fig_hm_dj}
    \end{subfigure}%
    \hfill
    \begin{subfigure}{.33\textwidth}
        \centering
        \includegraphics[width=\textwidth]{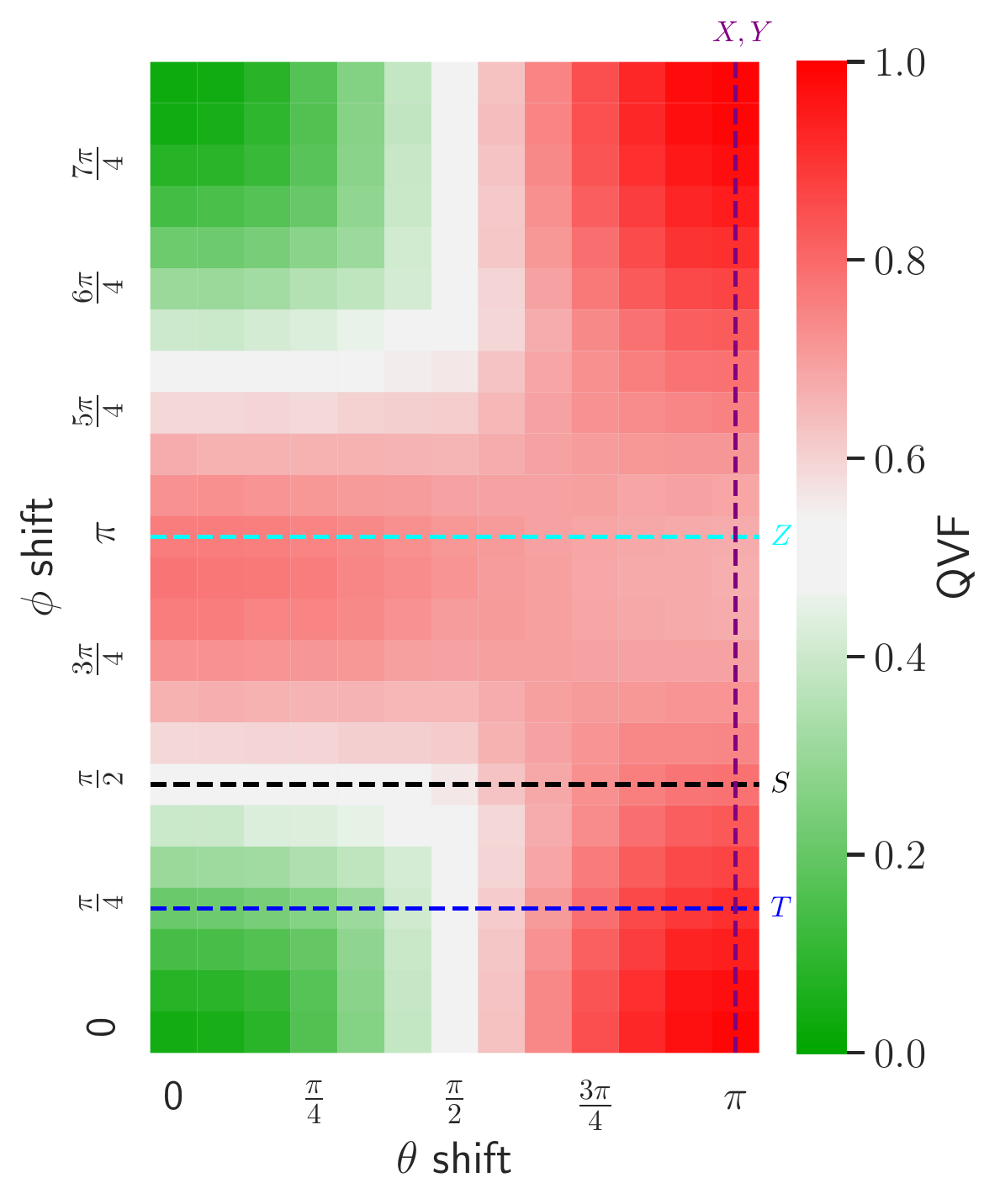}
        \caption{Grover.}
        \label{fig_hm_grover}
    \end{subfigure}%
    \\
        \begin{subfigure}{.3\textwidth}
    \captionsetup{justification=centering}
   	    \centering
        \includegraphics[width=\textwidth]{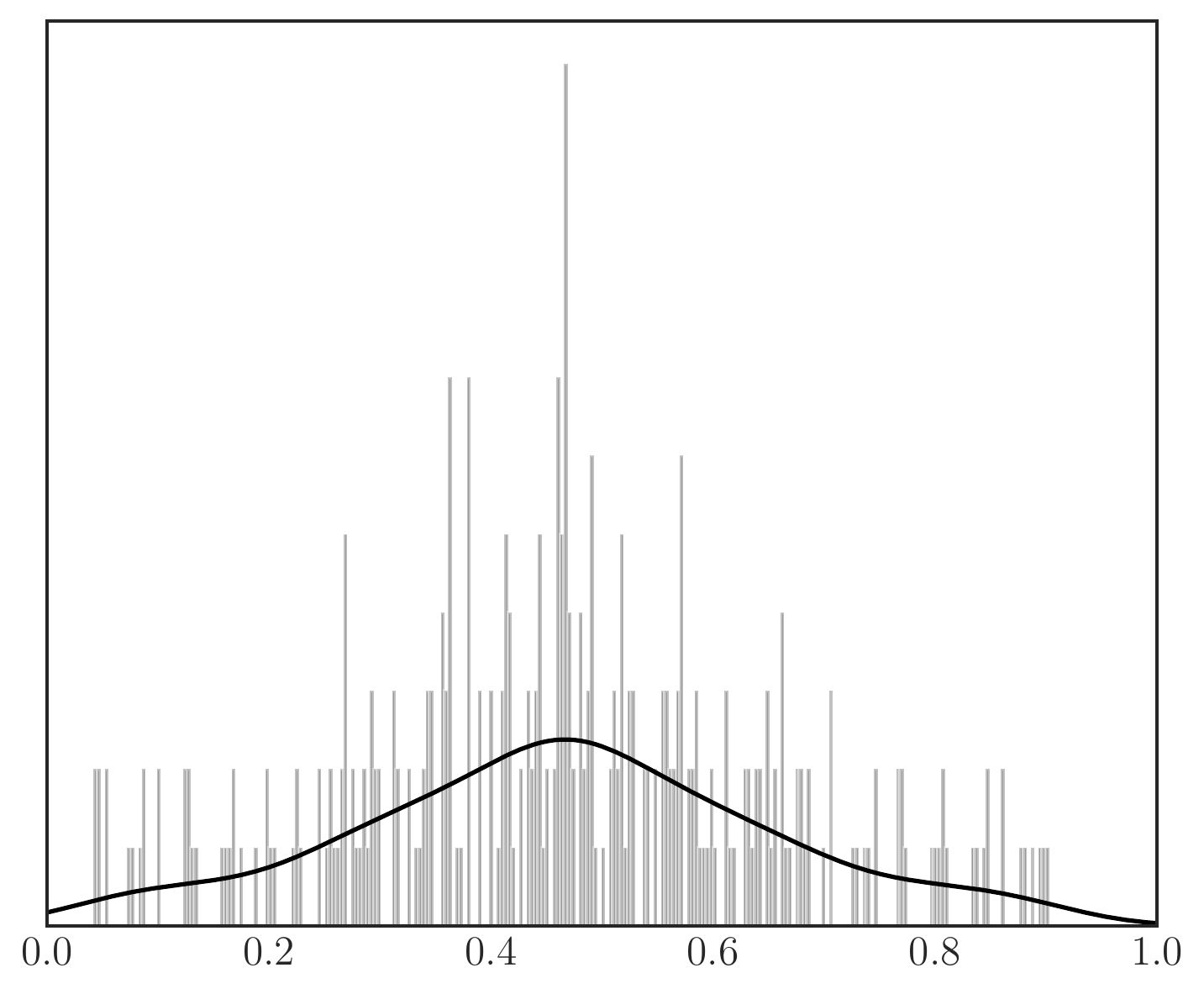}
        \caption{
            Bernstein-Vazirani\\Mean QVF=0.4656\\Stddev QVF=0.1884
        }
        \label{fig_hist_bv}
    \end{subfigure}%
    \hfill
    \begin{subfigure}{.3\textwidth}
    \captionsetup{justification=centering}
        \centering
        \includegraphics[width=\textwidth]{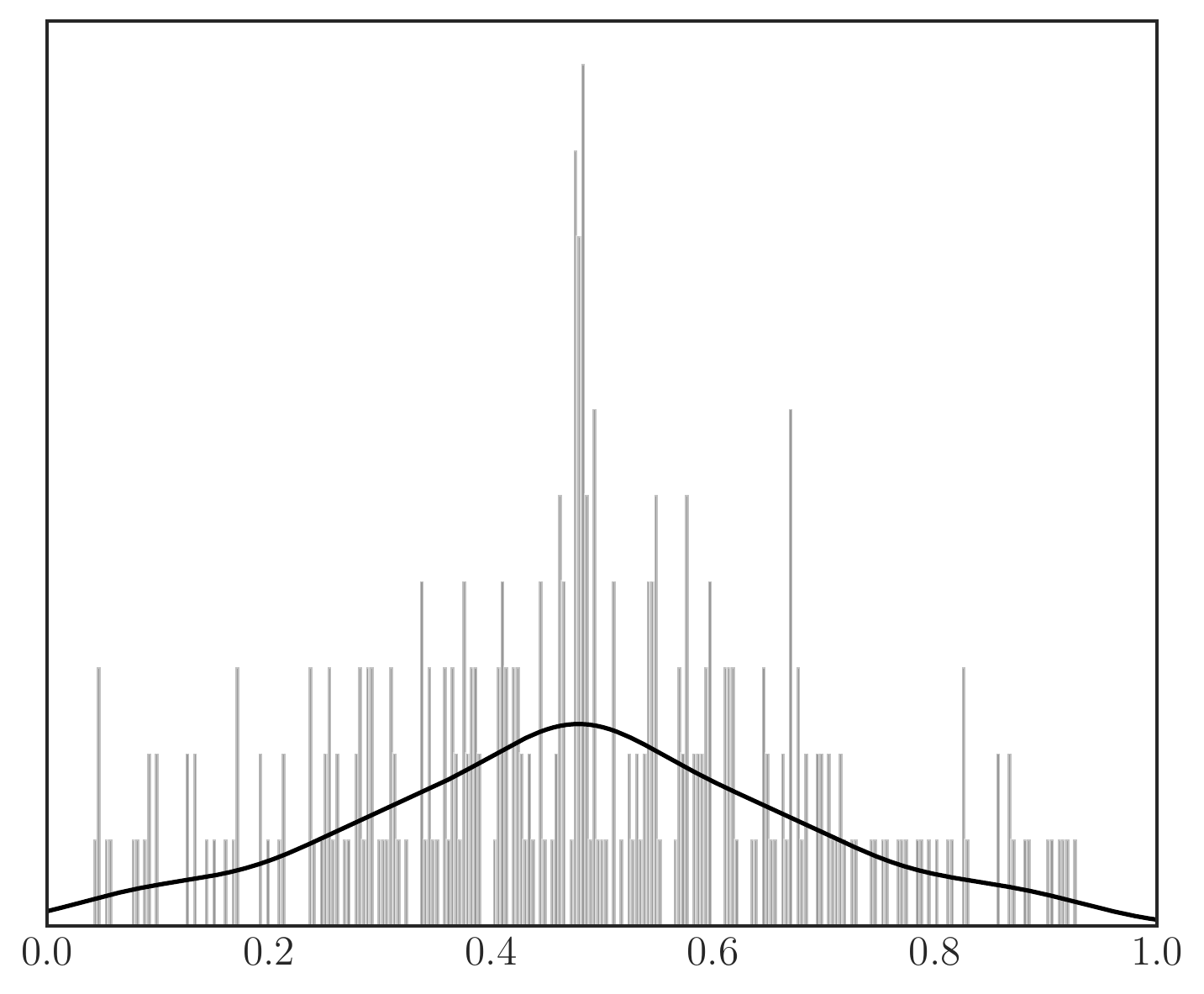}
        \caption{
            Deutsch-Josza\\Mean QVF=0.4791\\Stddev QVF=0.1944
        }
        \label{fig_hist_dj}
    \end{subfigure}%
    \hfill
    \begin{subfigure}{.3\textwidth}
    \captionsetup{justification=centering}
        \centering
        \includegraphics[width=\textwidth]{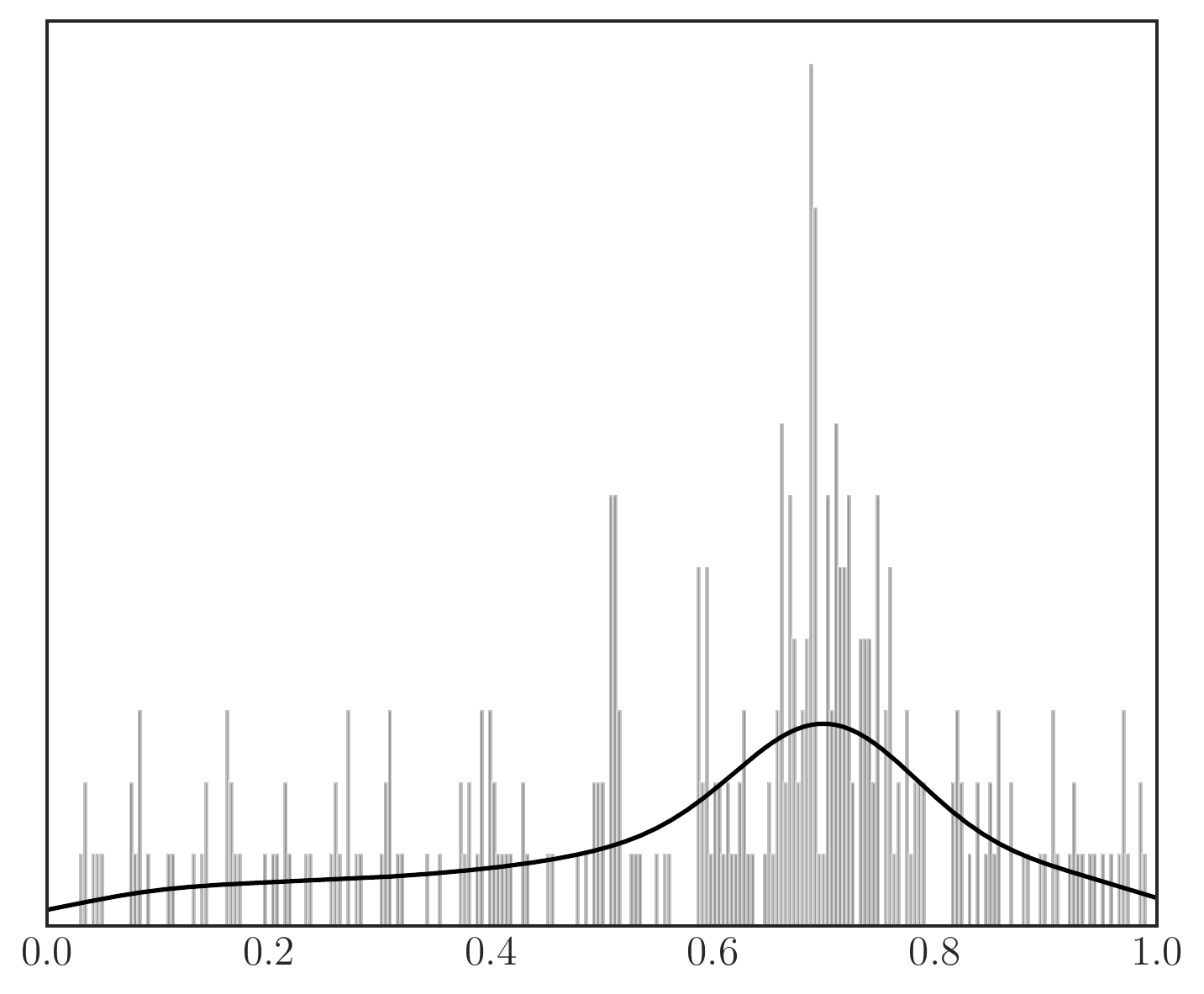}
        \caption{
            Grover\\Mean QVF=0.5975\\Stddev QVF=0.2289
        }
        \label{fig_hist_grover}
    \end{subfigure}%
    \caption{ \textbf{(a, b, c)} QVF heatmaps for the different circuits for different values of $\phi$ and/or $\theta$ shifts. The green color indicates a low QVF (the correct state can be confidently selected), the red color indicates a higher QVF (an incorrect output is more likely to be selected), and the white color indicates a dubious output (i.e., correct and incorrect states have about the same probability). The plot also shows dotted lines corresponding to the effect of common quantum gates ($X,Y,S,T,Z$) in order to provide a quick reference for the fault effect in the qubit.
    \textbf{(d, e, f)} Histograms of the QVF distribution of the three considered circuits.
    }
    \label{fig_heatmaps}
\end{figure*}

For the purpose of this study, we inject faults after each gate in the original circuit, simulating faults in each one of the circuit operations (see Figure~\ref{fig_fault_example}). We also inject only single qubit faults, which means that for each multiple qubit gate we generate multiple circuits with only one qubit at a time affected by the fault. For instance, a two-qubit CNOT gate will generate two distinct circuits with a fault on only one of the qubits each.
It is worth noting that our fault injector is also able to inject (correlated) multiple faults, as these events have already been observed in multiqubits circuits~\cite{muons2021}. However, we leave the multiple faults study for future work and in this paper, we limit the analysis to single faults since the scope of this work is to introduce the QVF metric and the fault injection framework, which are independent of specific fault models.

As the radiation-induced phase shifts can be of different magnitudes, we test several phase shifts. For each fault position we inject $312$ possible phase shifts (i.e., transient faults), and, for each fault, we execute the circuit $1,024$ times to obtain the probability distribution. Bernstein-Vazirani has 13 possible fault positions, thus requiring $4,056$ distinct faults and a total of $4,153,344$ executions. Grover and Deutsch-Jozsa have 18 fault positions each, which leads to $5,616$ distinct faults and $5,750,784$ executions. Thus, for the three circuits, we perform a fault injection campaign of $15,654,912$ total executions.
While this seems a large number of configurations, the time to generate the faulty circuits and the overhead of executing (or simulating) the faulty circuits are negligible. Once the phase shift is correlated with the particle impact, we can limit the injections to the relevant phase shifts. The original codes of circuits, data of faults injected, and output results are publicly available~\cite{REPO}.


Figure~\ref{fig_fault_example} illustrates the fault injection in a 2-qubit Grover circuit. The fault is injected in qubit $q_0$ after the first Hadamard gate with a $\theta$ phase shit of $\frac{\pi}{4}$. The probability distributions of the original and faulty circuit are plotted on the right. The correct state is $11$ and the $P(A)$ (i.e., the PST) is $0.937$ for the original circuit and $0.814$ for the faulty one. To compute the contrast in Equation~\ref{eq_contrast} and the QVF in Equation~\ref{eq_qvf}, we also need $P(B)$. For the original circuit $P(B)= P(10) = 0.040$, while for the faulty one $P(B) = P(01) = 0.150$. Then, the QVF is $0.04$ for the original and $0.16$ for the faulty circuit.

It is worth noting that the fault positions, as well as the fault model, can be easily modified to better correspond to reality as the understanding of quantum computer's reliability progresses.

\section{Experimental Results}
\label{sec_results}




In this Section, to illustrate and highlight the impact of QVF, we present the results obtained by injecting faults in three case-study quantum circuits: Bernstein-Vazirani (4 qubits), Deutsch-Josza (4 qubits), and Grover (2 qubits). 
We inject a fault in each qubit of each circuit, in all possible locations (i.e., after each gate).
As a radiation-induced fault can cause a $\phi$ and/or $\theta$ phase shift of various amplitudes, for each fault location we inject $\phi$ and/or $\theta$ angles with $15{^\circ}$ steps (i.e., $\frac{\pi}{12}$) for a range of $\phi=[0,\pi]$ and $\theta = [0,2\pi[$. Overall, we present data from more than $15,654,912$ injections.

\begin{figure*}[!ht]%
    \begin{subfigure}{.25\textwidth}
   	    \centering
        \includegraphics[width=\textwidth]{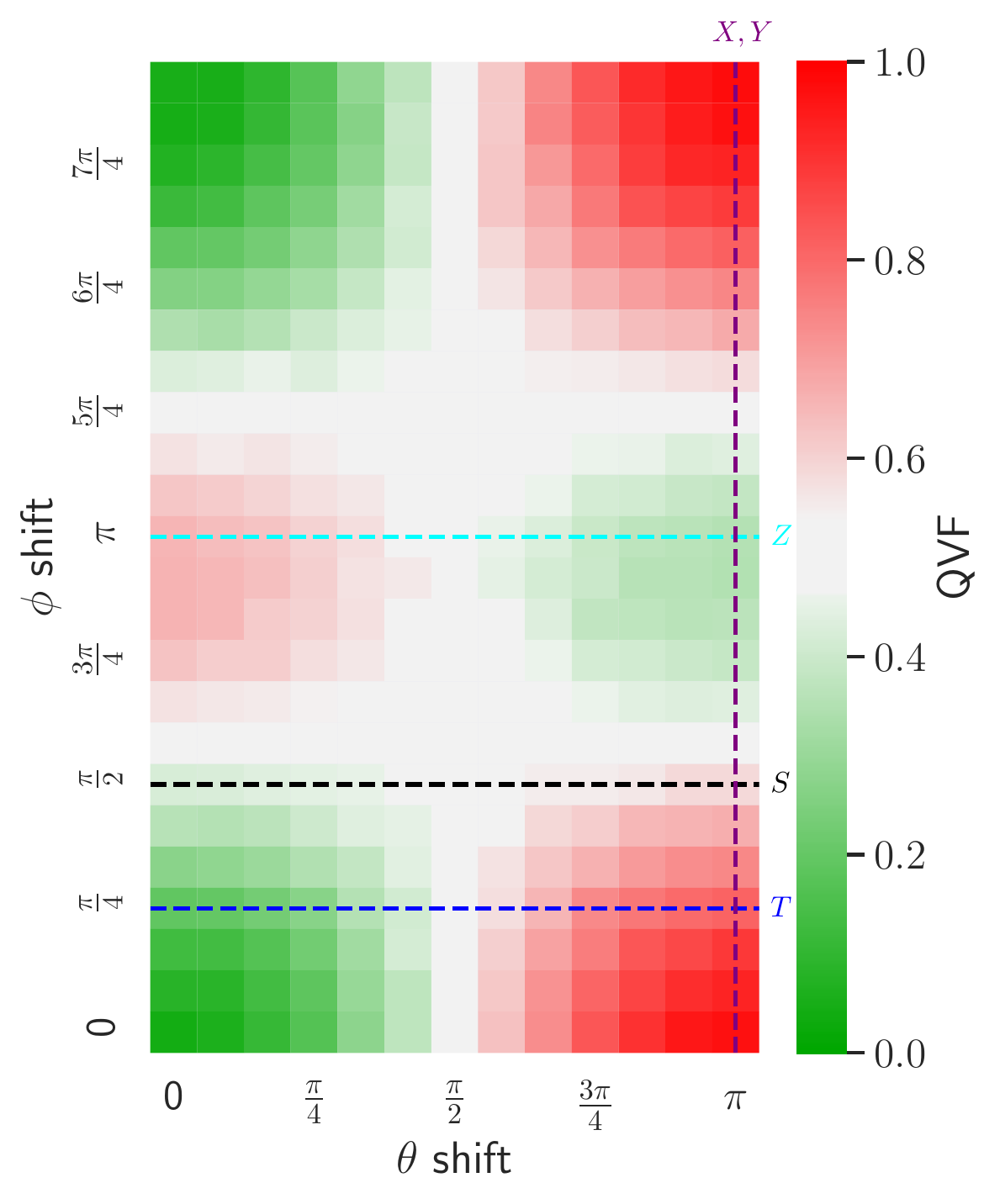}
        \caption{Qubit 0.}
        \label{fig_hm_bv_qb0}
    \end{subfigure}%
    \begin{subfigure}{.25\textwidth}
   	    \centering
        \includegraphics[width=\textwidth]{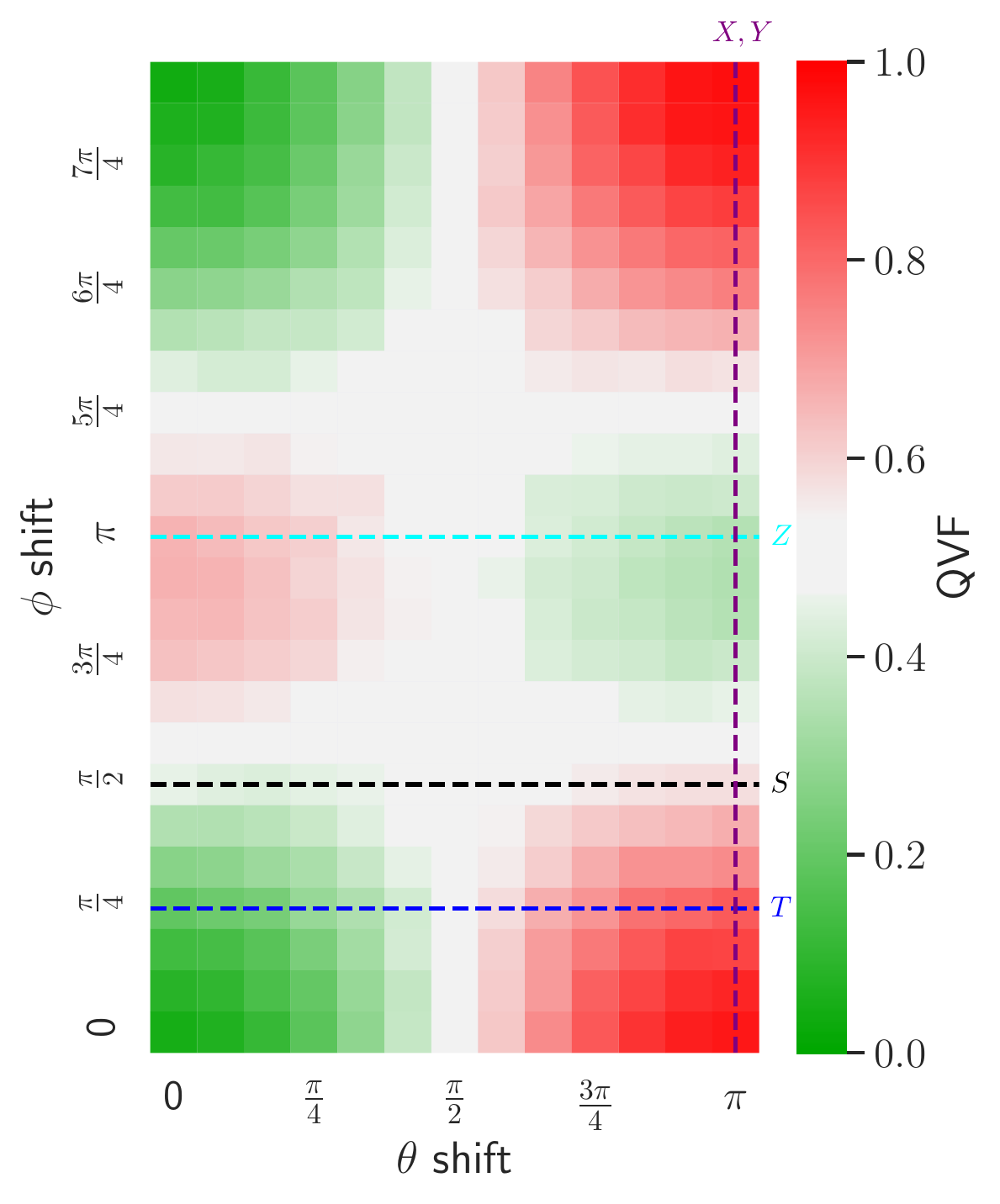}
        \caption{Qubit 1.}
        \label{fig_hm_bv_qb1}
    \end{subfigure}%
    \begin{subfigure}{.25\textwidth}
   	    \centering
        \includegraphics[width=\textwidth]{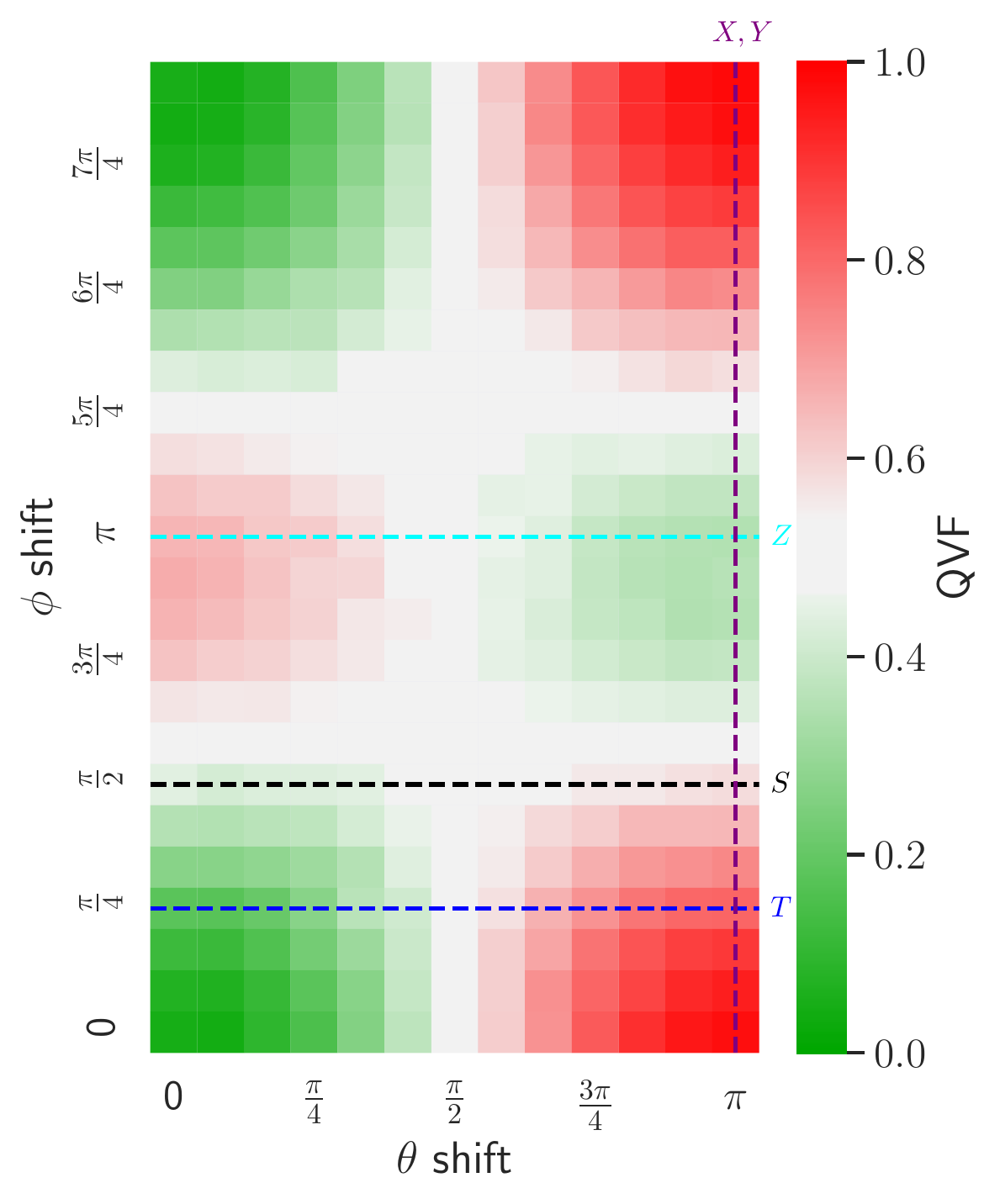}
        \caption{Qubit 2.}
        \label{fig_hm_bv_qb2}
    \end{subfigure}%
    \begin{subfigure}{.25\textwidth}
   	    \centering
        \includegraphics[width=\textwidth]{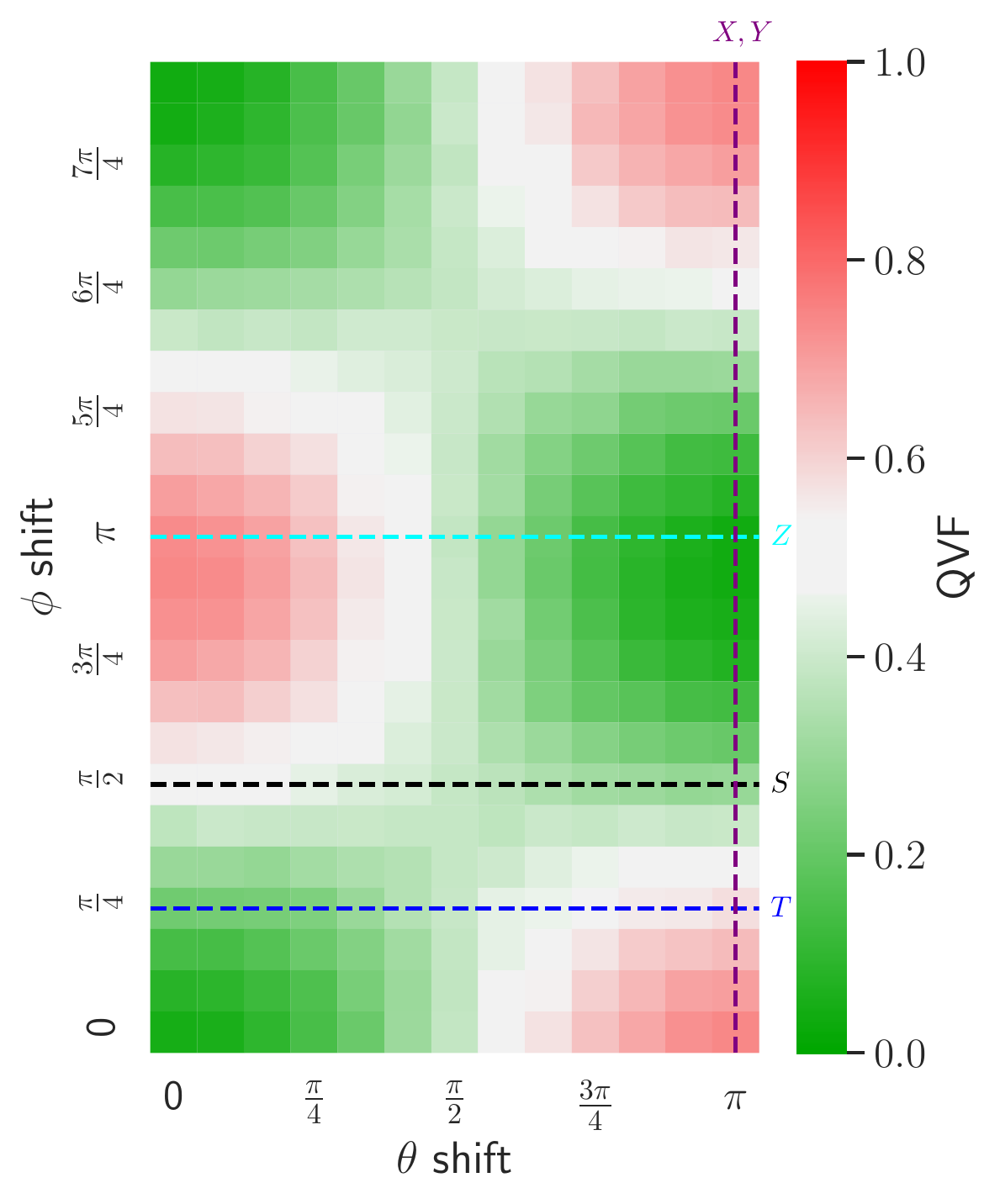}
        \caption{Qubit 3.}
        \label{fig_hm_bv_qb3}
    \end{subfigure}%
    \caption{QVF, per qubit, for Bernstein-Vazirani circuit for different values of $\phi$ and $\theta$. The green color indicates that the correct state can be confidently selected, the red color an incorrect one will be selected, and the white color produces a dubious output (i.e., correct and incorrect states have about the same probability).}
    \label{fig_heatmaps_bv}
\end{figure*}

Unlike classical computing, we need to consider a great number of possible configurations for each fault location, and each injection can have various impacts on the circuit output.
We plot different graphs to visually appreciate the fault impact in a quantum circuit output.
We start the QVF analysis by showing, in Figure~\ref{fig_heatmaps}, the heatmap of the QVF for the considered quantum circuits.
We plot, for each circuit, the QVF computed for each fault (i.e., the injected $\phi$ and/or $\theta$ shifts).
Each spot $(\phi,\theta)$ represents the QVF mean for all possible fault injections (qubit and position inside a qubit) for that specific $(\phi,\theta)$ phase shift. To have a more direct reference of the effect of faults in the qubit state, in Figure~\ref{fig_heatmaps} we also superposed colored lines corresponding to the phase shifts that would be imposed by common quantum gates ($X,Y,S,T,Z$). 

Green colors in Figure~\ref{fig_heatmaps} (QVF $< 0.45$) mean the circuit still produces the correct output as the most likely one. 
White colors ($0.45 <$ QVF $< 0.55$) means that the fault causes the output to be dubious (i.e., the correct output cannot be confidently selected). Finally, red colors (QVF $> 0.55$) means that the fault effect is so high to produce an incorrect output as the most likely, leading to errors.
These colors are used to ease the visualization and can be changed when an acceptable threshold for QVF is defined. 

Let us consider, in Figure~\ref{fig_heatmaps}, the {$(\phi=0, \theta=0)$} spot.
This square represents the QVF computed for the fault-free, yet noisy, execution of the circuits.
The value for this square is not solid green (i.e., QVF $= 0$), since the circuit itself has its own imperfections due to noise.
Interestingly, we found that, in some rare cases ($\sim$3\%), faults actually improve the circuit QVF compared to the fault-free (but noisy) execution.
This happens because the injected fault compensates for the noise effect, making the output state distribution closer to the ideal case. The fact that a fault improves the output quality should not surprise, as it has been observed in probabilistic classical - such as artificial neural networks (despite with much lower probability)~\cite{tr2019} - and quantum computation applications \cite{ayanzadeh2021equal}.

\begin{figure*}[!ht]%
    \begin{subfigure}{.33\textwidth}
   	    \centering
        \includegraphics[width=\textwidth]{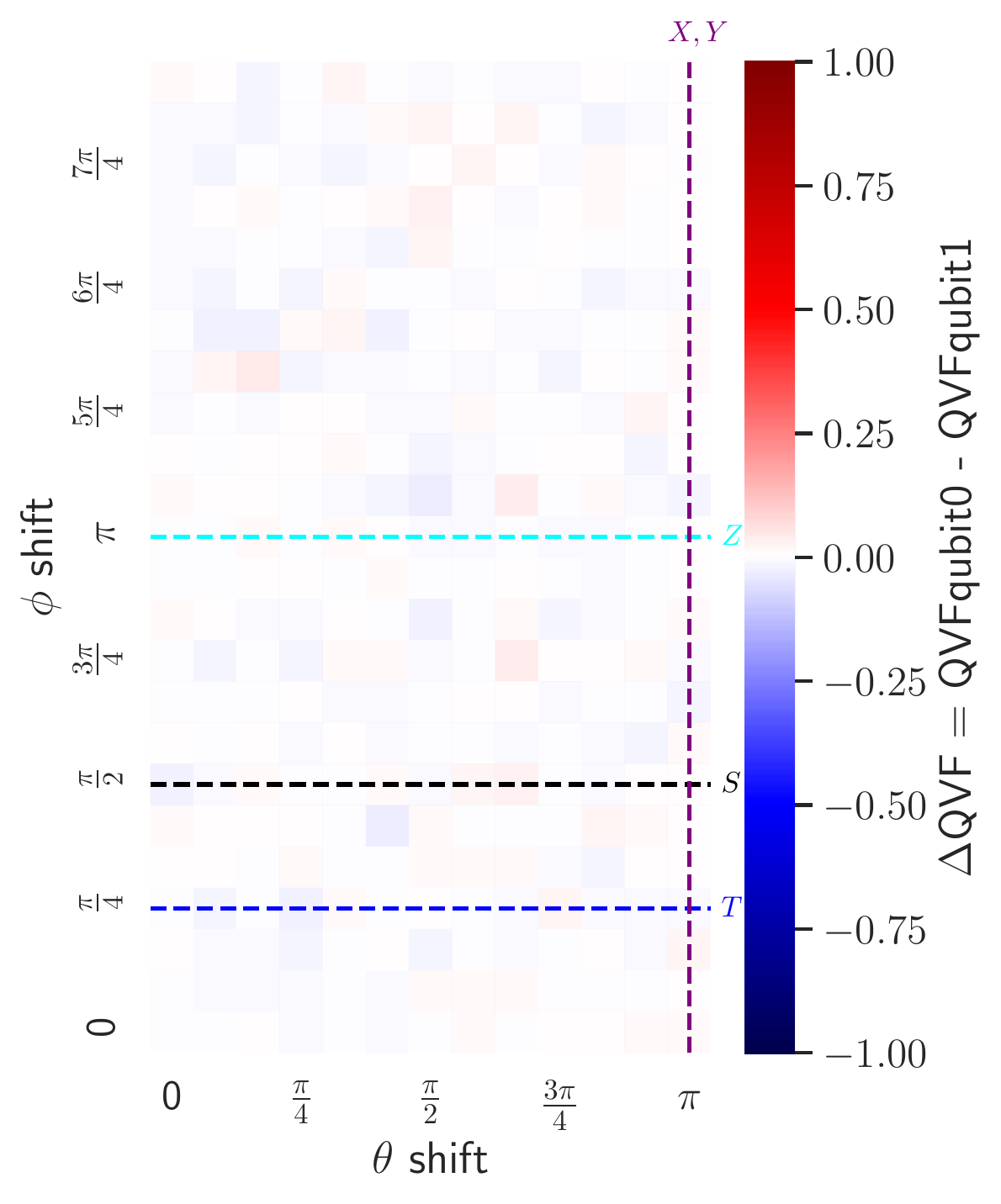}
        \caption{$\Delta$ Qubit0 - Qubit1.}
        \label{fig_hm_bv_qb0_qb1}
    \end{subfigure}%
    \begin{subfigure}{.33\textwidth}
   	    \centering
        \includegraphics[width=\textwidth]{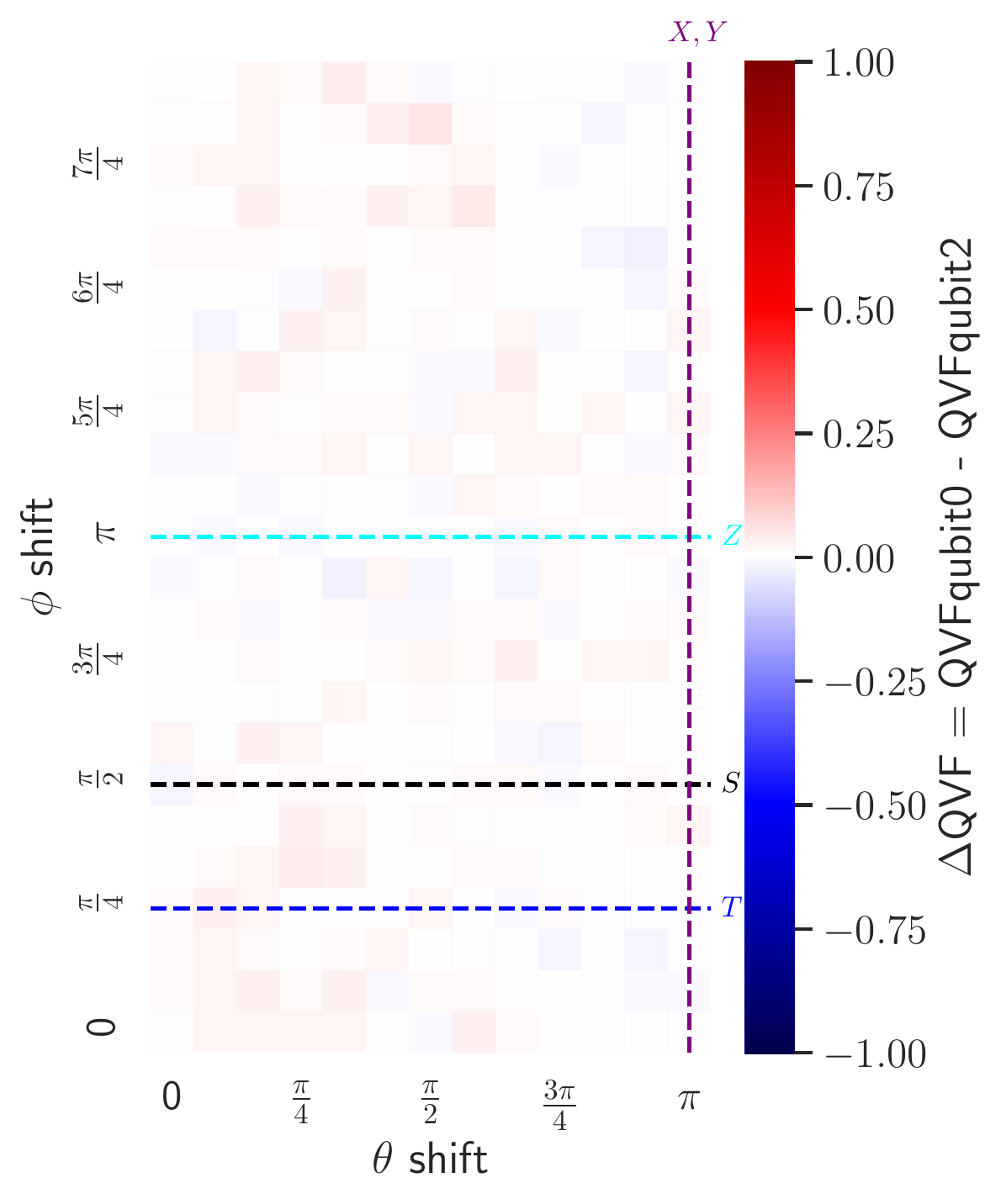}
        \caption{$\Delta$ Qubit0 - Qubit2.}
        \label{fig_hm_bv_qb0_qb2}
    \end{subfigure}%
    \begin{subfigure}{.33\textwidth}
   	    \centering
        \includegraphics[width=\textwidth]{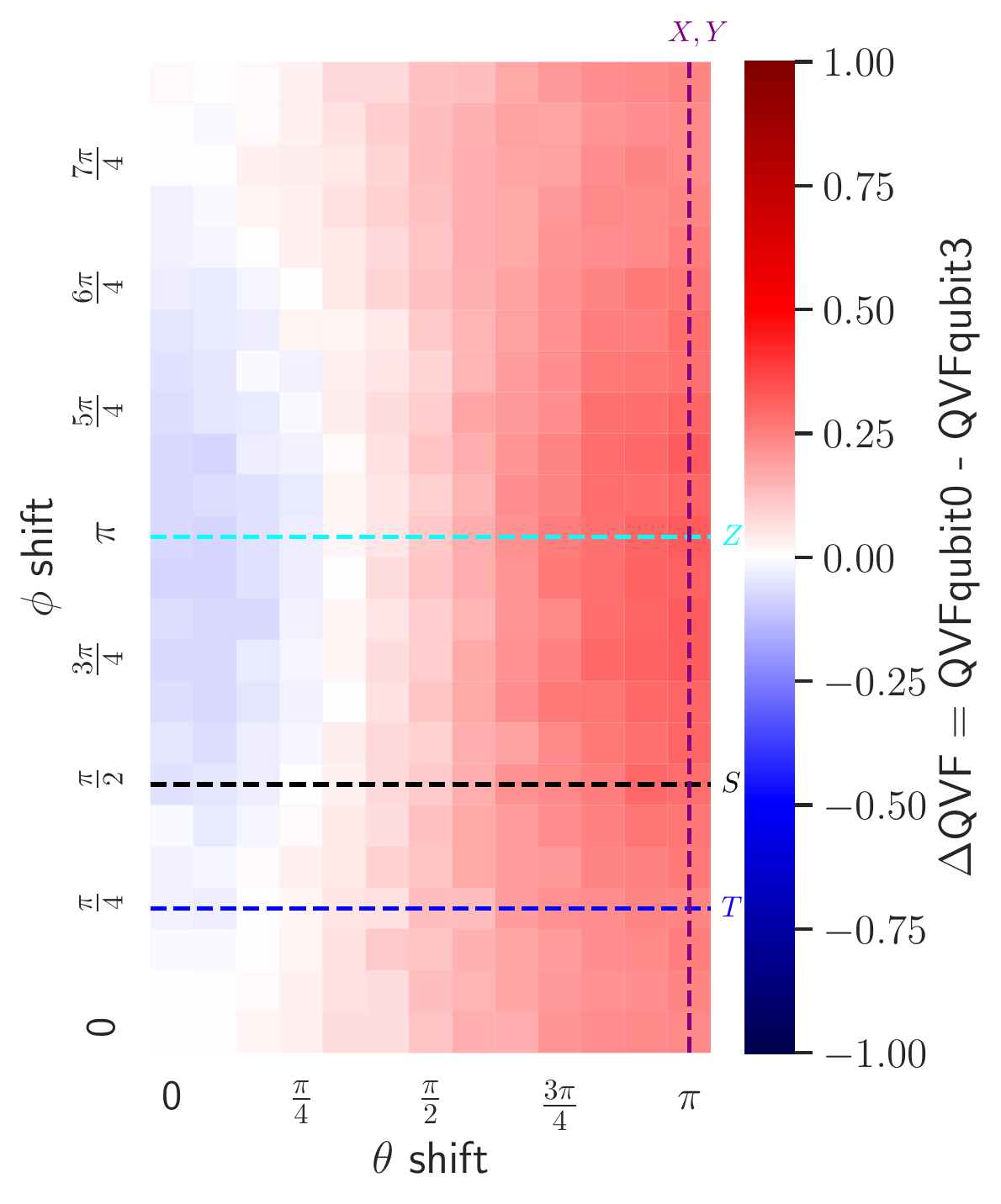}
        \caption{$\Delta$ Qubit0 - Qubit3.}
        \label{fig_hm_bv_qb0_qb3}
    \end{subfigure}%
    \caption{ $\Delta QVF$ for Bernstein-Vazirani circuit between different qubits.}
    \label{fig_delta_bv}
\end{figure*}

\begin{figure*}[t]%
    \begin{subfigure}{.33\textwidth}
   	    \centering
        \includegraphics[width=\textwidth]{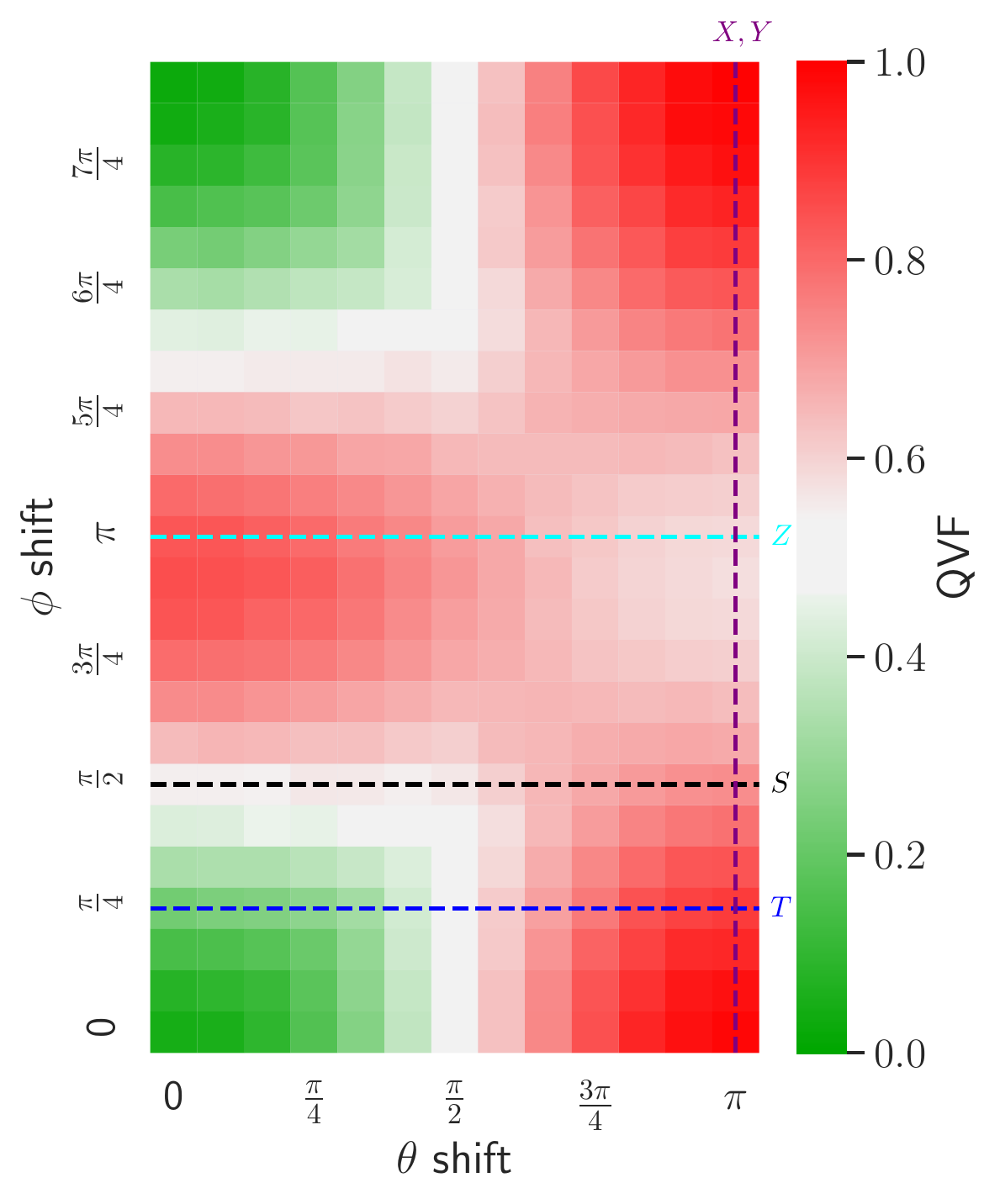}
        \caption{Qubit 0.}
        \label{fig_hm_grover_qb0}
    \end{subfigure}%
    \begin{subfigure}{.33\textwidth}
   	    \centering
        \includegraphics[width=\textwidth]{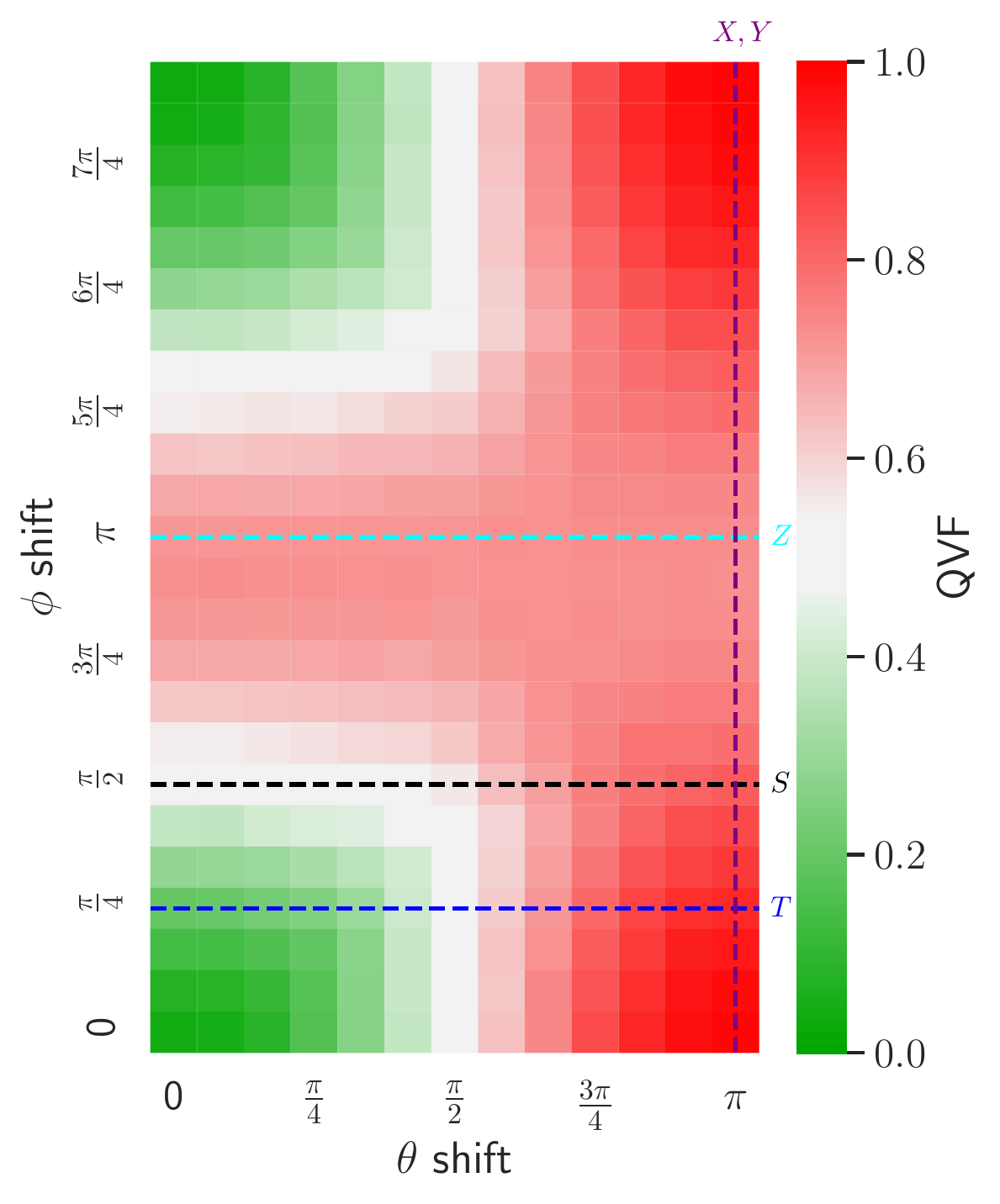}
        \caption{Qubit 1.}
        \label{fig_hm_grover_qb1}
    \end{subfigure}%
    \begin{subfigure}{.33\textwidth}
   	    \centering
        \includegraphics[width=\textwidth]{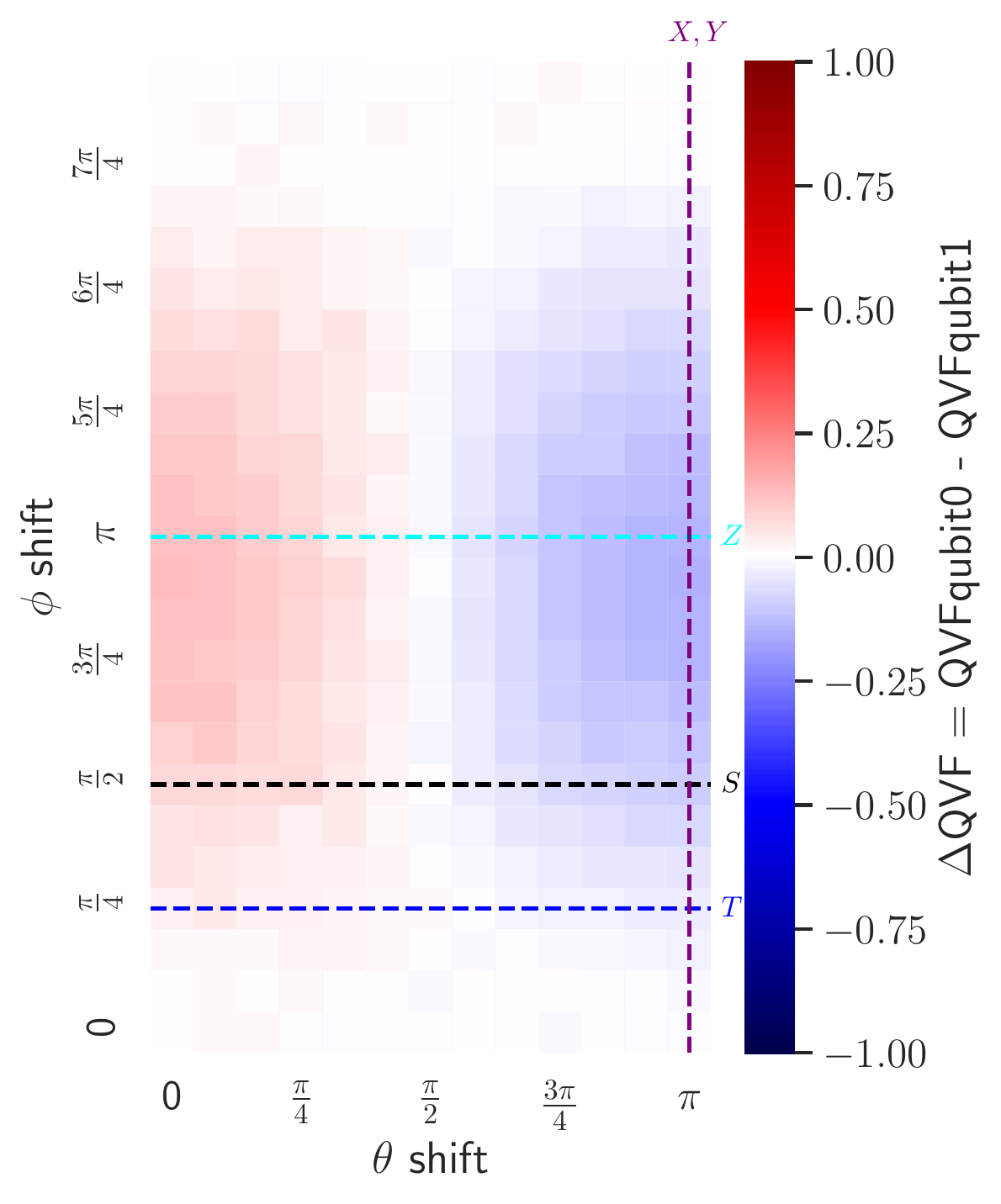}
        \caption{$\Delta$ Qubit 0 - Qubit 1.}
        \label{fig_hm_grover_qb0_qb1}
    \end{subfigure}%
    \caption{$QVF$ for Grover circuit for each qubit and the $\Delta QVF$ between them.}
    \label{fig_delta_grover}
\end{figure*}

By fixing the injected $\phi$ shift to 0 (we are keeping for $\phi$ the same value of the noisy, but fault-free, execution) and moving to the right in Figure~\ref{fig_heatmaps} we can study the effect of injecting a fault with increasing $\theta$ from 0 to $\pi$ (without shifting $\phi$). In other words, we are gradually modifying the 0-1 probability in the qubit without changing the $\phi$ phase. As shown in Figure~\ref{fig_heatmaps}, increasing $\theta$ keeping the fault-free $\phi$ increases the QVF and, thus, makes the circuit highly unreliable. 
It is worth noting that a $\frac{\pi}{2}$ shift in $\theta$ is the point where the output becomes dubious, and that is exactly the angle ($90^\circ$) where the direction starts to flip in the Bloch sphere. For shifts greater than $\frac{\pi}{2}$ the 0-1 probability of the qubit is effectively changed and an incorrect state becomes the most likely one. 

A similar trend is observed for $\phi$ when $\theta$ is not shifted, however, the QVF is not as severely impacted as for $\theta$, resulting in lower QVF values but still higher than 0.55 (red colors). Thus, a shift in $\theta$ (i.e., a shift in the 0-1 state probability) is indeed more critical than a shift in $\phi$. While this seems not surprising, please note that for Bernstein-Vazirani and Deutsch-Jozsa, the combination of a $\theta$ and $\phi$ shift (e.g., $(\phi=\pi, \theta=\pi)$ has a beneficial effect on the QVF. Thus, this combination seems to compensate for the shift resulting in still acceptable QVFs (green colors). In contrast, Grover has a different response to the combination of a $\theta$ and $\phi$ shift, and it is not sufficient to produce acceptable QVFs. For instance, QVF mean for $(\phi=\pi, \theta=\pi)$ is $0.254224$, $0.249150$, and $0.666607$ for Bernstein-Vazirani, Deutsch-Jozsa, and Grover, respectively.


An additional interesting insight is that Figure~\ref{fig_heatmaps} is almost symmetric on $\phi$ with respect to $\pi$. This is justified as moving towards $2\pi$ on the Bloch sphere as we pass $\pi$ we move closer to the original position of $\phi$.

Figure~\ref{fig_heatmaps} can also be used to have a first comparison of the reliability of circuits. The higher the number of red spots, the higher the number of faults that can corrupt the circuit output. 
Figure~\ref{fig_heatmaps} (d, e, f) are in fact depicting the histograms of QVF for the three algorithms, showing also the QVF mean value and standard deviation. Grover algorithm presents, in the QVF map of all faults, a mean value of $0.59$, which means that the circuit grants more probability to the wrong result than to the right one ($0.5$ is the automatic threshold). The other two algorithms are much better, with both a lower mean value (lower than $0.5$) and a reduced standard deviation.
Histograms can provide a method that does not require human intervention and that could be applied to a large number of random circuits and/or specific faults.
Histogram plotting and other image processing techniques can be applied to the whole image or on a subsection of it.


\begin{figure*}[t]%
    \begin{subfigure}{.33\textwidth}
   	    \centering
        \includegraphics[width=\textwidth]{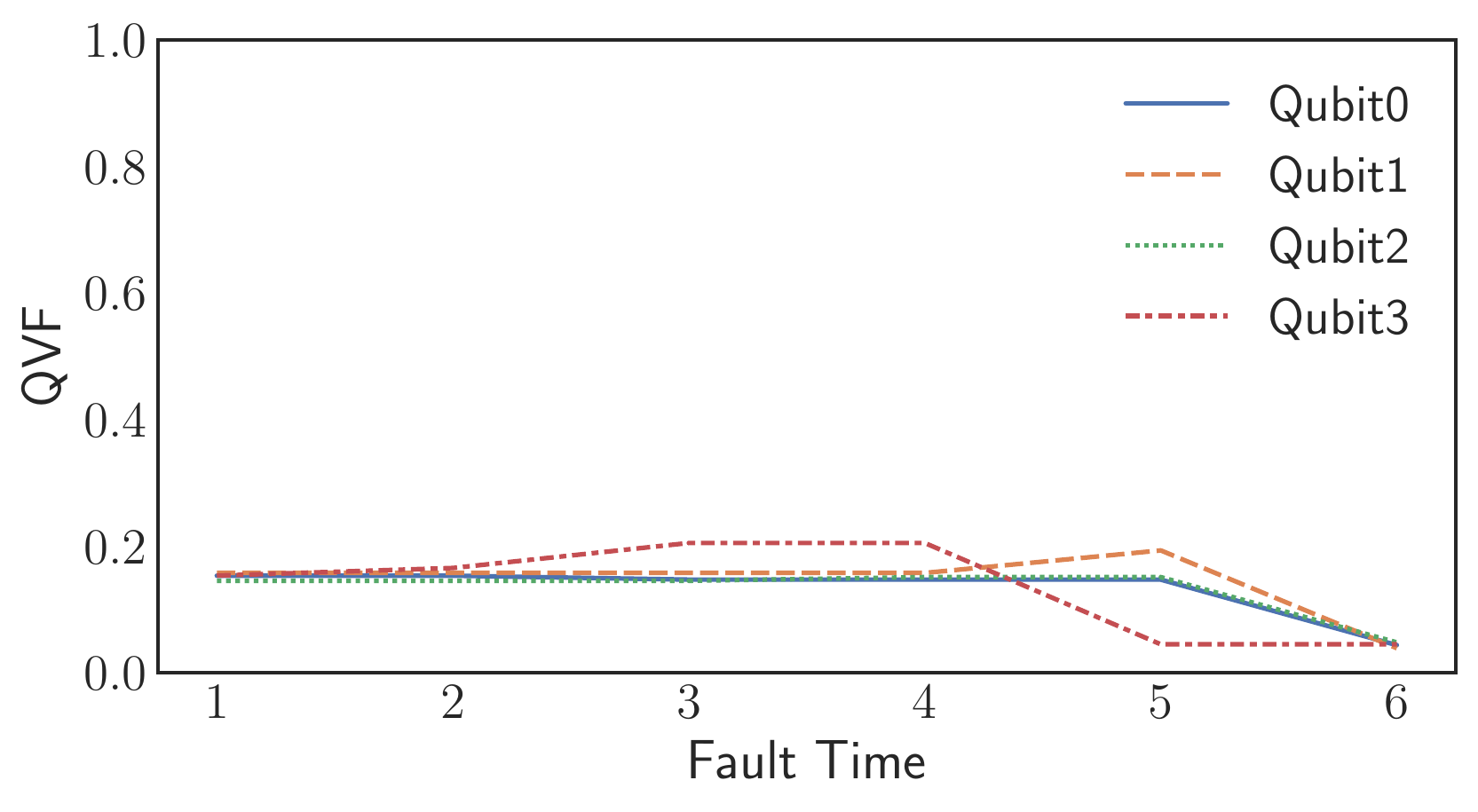}
        \caption{Bernstein-Vazirani.}
        \label{fig_ft_tgate_bv}
    \end{subfigure}%
    \begin{subfigure}{.33\textwidth}
        \centering
        \includegraphics[width=\textwidth]{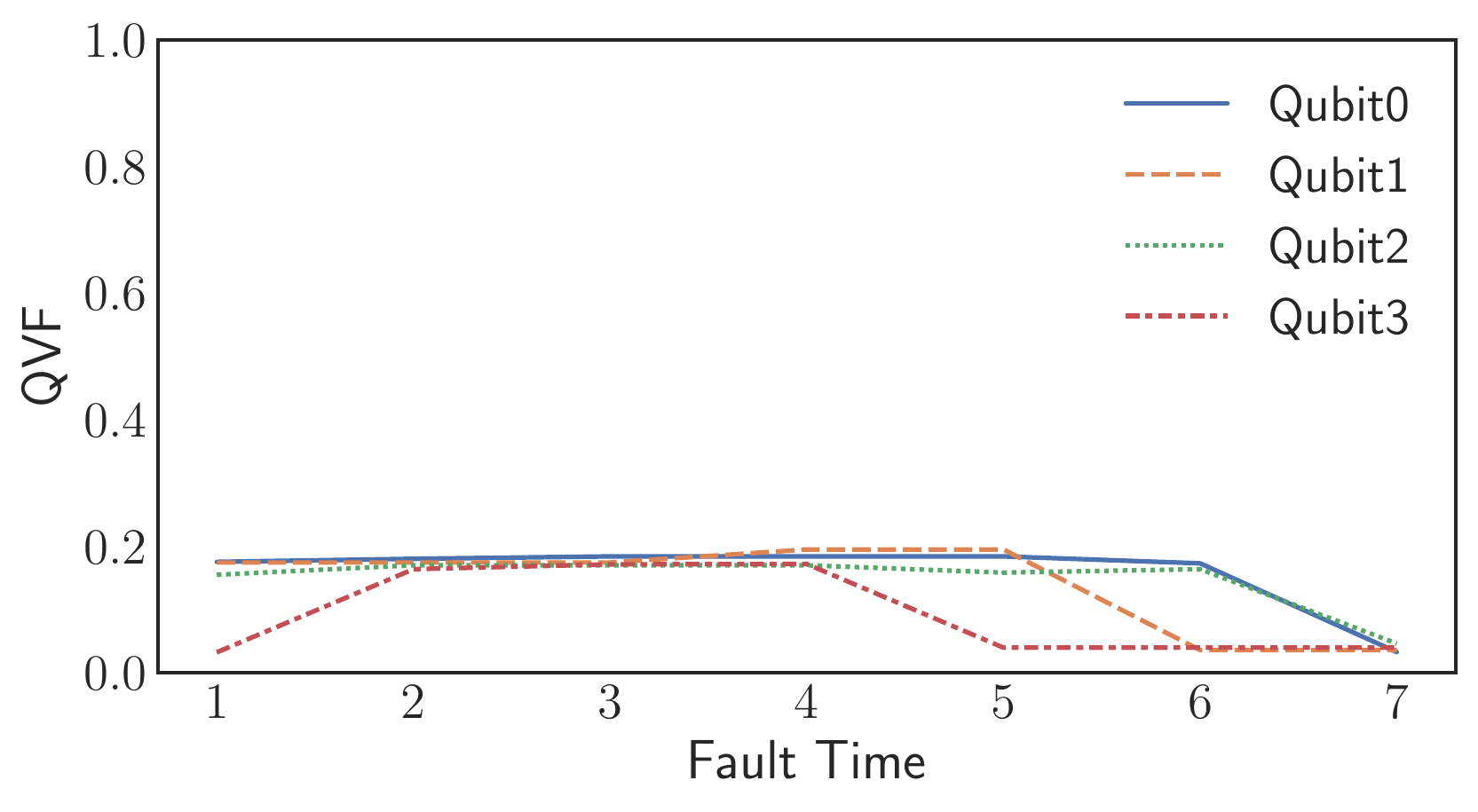}
        \caption{Deutsch-Jozsa.}
        \label{fig_ft_tgate_dj}
    \end{subfigure}%
    \begin{subfigure}{.33\textwidth}
        \centering
        \includegraphics[width=\textwidth]{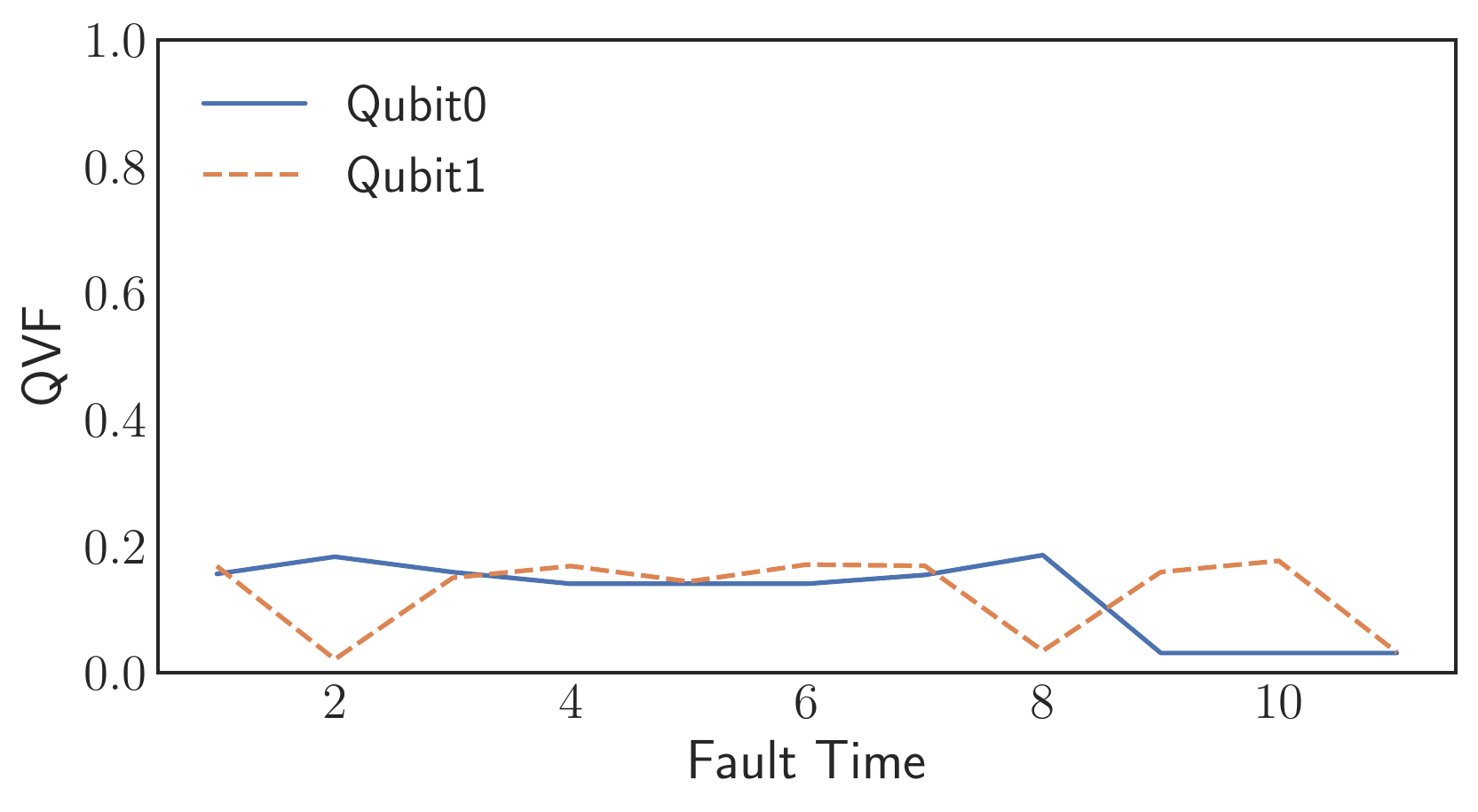}
        \caption{Grover.}
        \label{fig_ft_tgate_grover}
    \end{subfigure}%
    \caption{Example of QVF time dependency obtained injecting the fault $\theta=0$ and $\phi=\frac{\pi}{2}$ (T gate).
    }
    \label{fig_ft_tgate}
\end{figure*}

\begin{figure*}[t]%
    \begin{subfigure}{.33\textwidth}
   	    \centering
        \includegraphics[width=\textwidth]{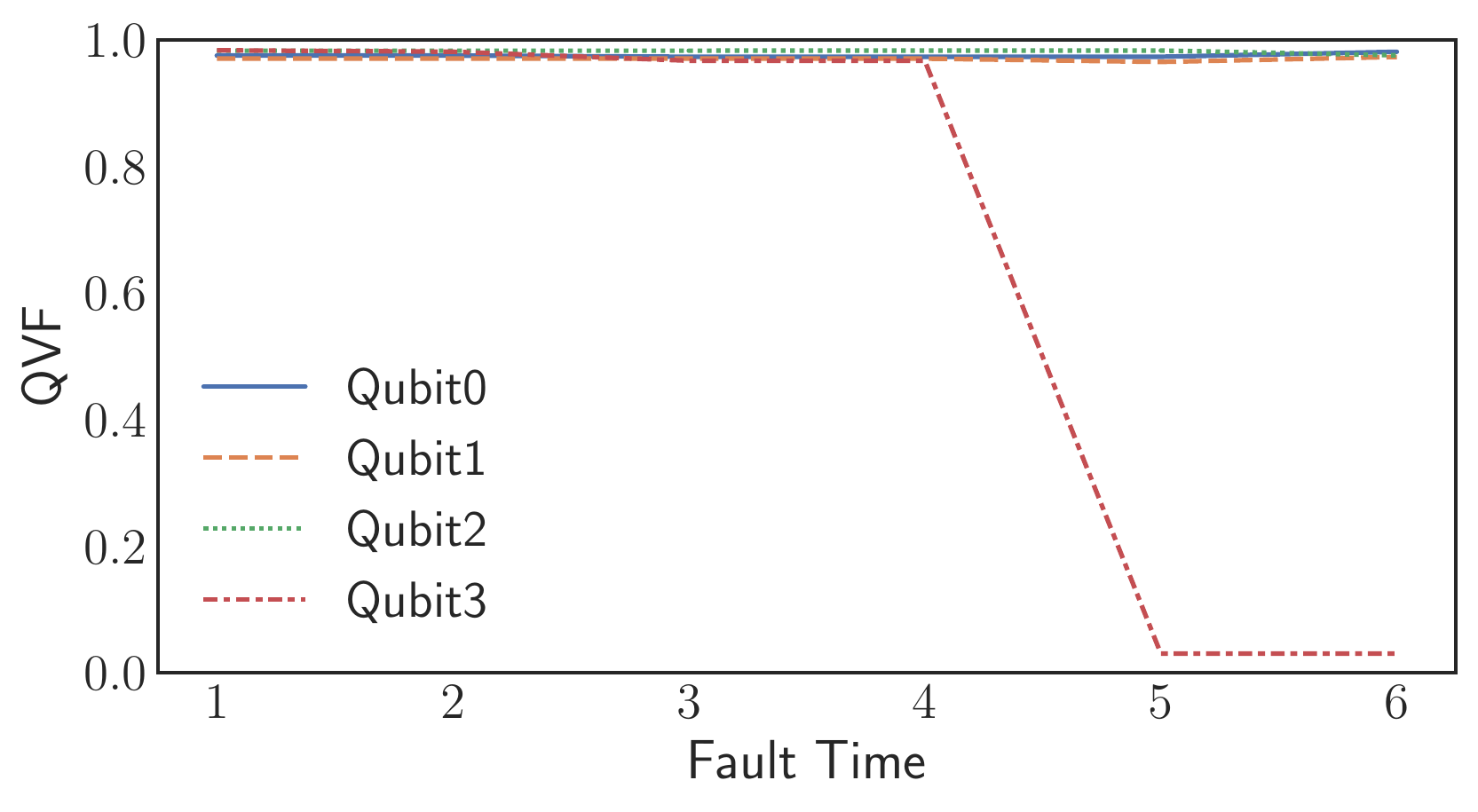}
        \caption{Bernstein-Vazirani.}
        \label{fig_ft_ygate_bv}
    \end{subfigure}%
    \begin{subfigure}{.33\textwidth}
        \centering
        \includegraphics[width=\textwidth]{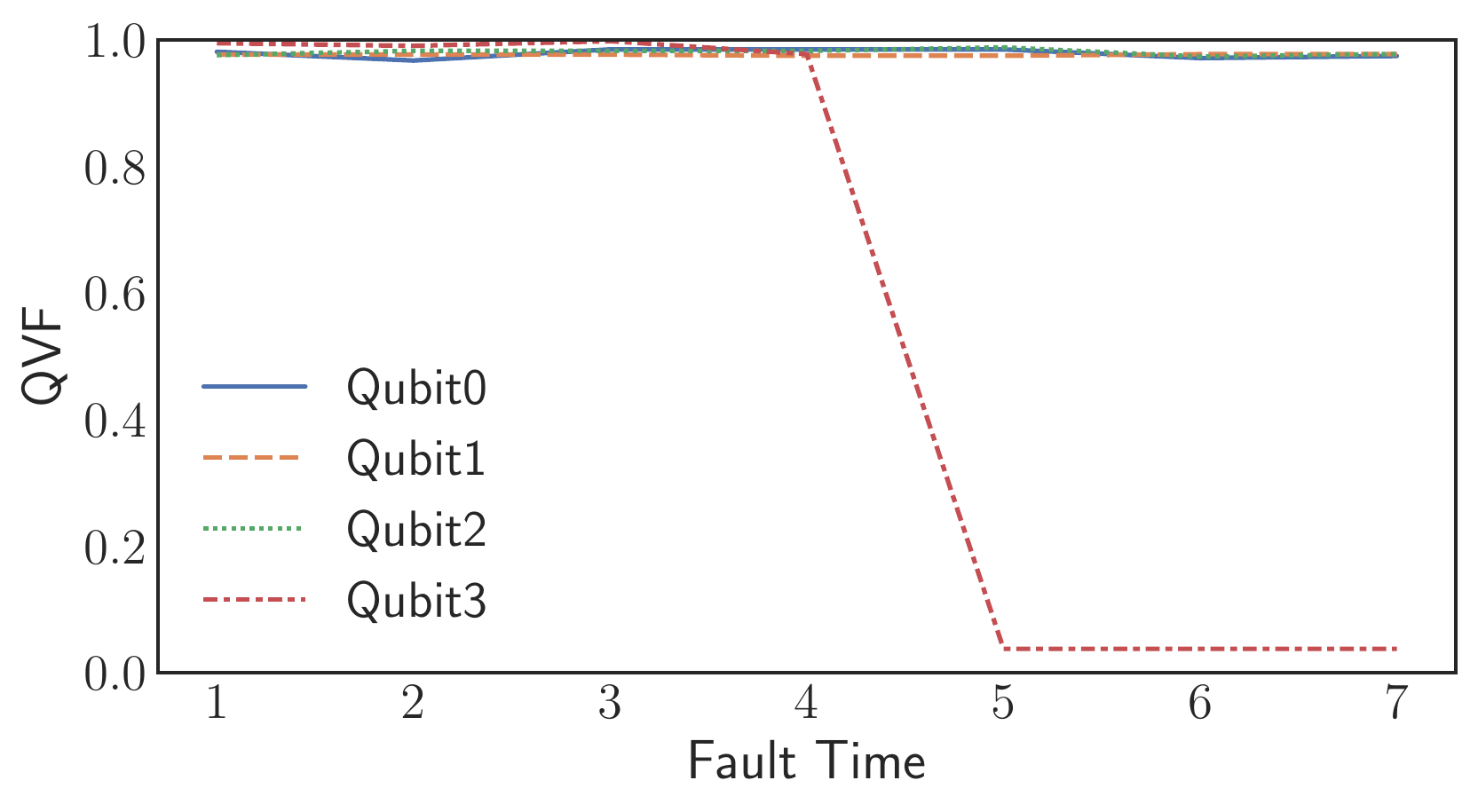}
        \caption{Deutsch-Jozsa.}
        \label{fig_ft_ygate_dj}
    \end{subfigure}%
    \begin{subfigure}{.33\textwidth}
        \centering
        \includegraphics[width=\textwidth]{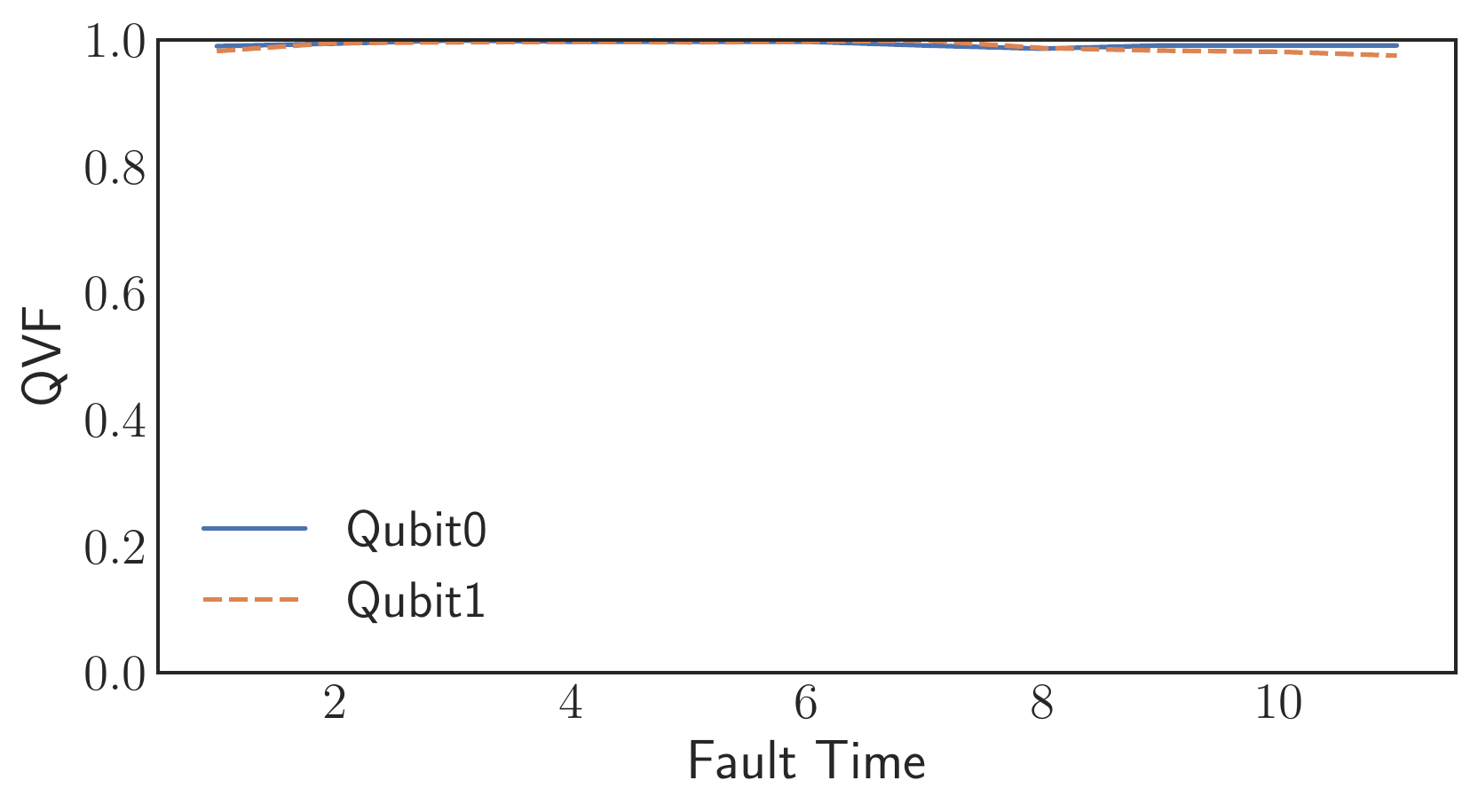}
        \caption{Grover.}
        \label{fig_ft_ygate_grover}
    \end{subfigure}%
    \caption{Example of QVF time dependency obtained injecting the fault $\theta=\pi$ and $\phi=0$ (Y gate).
    }
    \label{fig_ft_ygate}
\end{figure*}

Additionally, Figure~\ref{fig_heatmaps} gives an indication of the faults that are more critical for a given circuit. For instance, any shift with $\theta > \frac{\pi}{2}$ and/or $\frac{\pi}{2} < \phi < \frac{5\pi}{4}$ is critical for Grover. However, this is not the case for Bernstein-Vazirani nor Deutsch-Jozsa, indicating that the fault criticality is circuit-dependent and cannot be assumed a priori. 
We acknowledge the lack of an experimental evaluation of the phase shift amplitude resulting from the particles hit. Assuming, as for CMOS technology, that the position and energy of the impinging particles are stochastic, in Figure~\ref{fig_heatmaps} we assume each phase shift to be equally probable.
As we improve the understanding of the fault model and the probability of a specific $(\phi, \theta)$ shift to occur, we can normalize the QVF for each $(\phi, \theta)$ probability and provide a more realistic comparison. 


An insight we can derive from the QVF evaluation is the identification of the qubit(s) in a circuit that is more likely, if corrupted, to impact the output correctness.
This is fundamental information as it allows to focus the design and implementation of extra fault tolerance solutions where they are more needed.


To better understand the impact of a fault depending on the faulty qubit, we plot, in Figure~\ref{fig_heatmaps_bv}, the heatmap of QVF for each one of the four qubits of the Bernstein-Vazirani circuit (we do not plot the heatmap for the qubits of the other circuits for lack of space). 
These plots show that there are areas ($\phi$ and/or $\theta$ shift injections) that are more critical for circuit correctness than others.
The plot for qubit 3, being lighter among the four qubits, implies that it is less likely for a fault in qubit 3 to impact the overall behavior of the circuit. In fact, the QVF mean for qubits 0 to 3 in $(\phi=\pi, \theta=\pi)$ is, respectively, $0.348350$, $0.347449$, $0.357679$, and $0.036119$.
This is justified by the fact that qubit 3 in Bernstein-Vazirani acts as an ancilla qubit (i.e., auxiliary qubit). As qubit 3 is not directly measured at the output, it has a lower QVF. This is confirmed in the discussion on the time dependence of the QVF that we present next (see Figures~\ref{fig_ft_tgate} and~\ref{fig_ft_ygate}). 

We can better understand the different vulnerabilities of qubits from the heatmaps in Figure~\ref{fig_delta_bv}, in which we plot, for the Bernstein-Vazirani circuit, the $\Delta$QVF, i.e. the difference between the QVF of each pair of qubits. Please note that values higher than zero (red colors) indicate that the former qubit has a higher QVF, and thus a higher impact on the circuit output than the latter. $\Delta$QVF lower than zero (blue colors) indicates the opposite, with the former qubit having a lower QVF, and thus better reliability than the latter.
There is not much difference between qubit 0 and qubit 1 (Fig.\ref{fig_hm_bv_qb0_qb1}) or between qubit 0 and qubit 2 (Fig.\ref{fig_hm_bv_qb0_qb2}). The maximum $\Delta$QVF for both cases is $0.048622$ and the minimum $-0.034$. However, there is definitely an appreciable difference between qubit 0 and qubit 3 (Fig.\ref{fig_hm_bv_qb0_qb3}), for which the maximum $\Delta$QVF is $0.317$ and, for $\theta \geq \frac{\pi}{2}$, all values are higher than $0.127$.
In this case, qubit 0 performs worse than qubit 3 with an overall $\Delta$QVF mean of $0.113$, hence the prevailing red colors. Thus, as observed before, qubit 3 acts as an ancilla qubit and has a lower impact on the circuit.

Figure~\ref{fig_delta_grover} shows the QVF for the two qubits in the Grover circuit and the delta between them.
From these plots, it is possible to see that the two qubits have a mirrored performance with respect to $\theta=\frac{\pi}{2}$ and centered on $\phi=\pi$.
This means that, starting from $\phi=\pi$ and $\theta=\frac{\pi}{2}$ and moving to the left, one qubit performs worse and the other better, while if we move to the right we have the opposite trend.

A final and interesting evaluation allowed by QVF and our fault injection framework regards the time sensitivity of qubits.
We can identify if the effect of a fault in a qubit depends on the moment (or position in the circuit) in which the fault occurs. As discussed in Section~\ref{sub_radiation}, heavy particles hits are stochastic and can happen at any moment of the circuit execution. We aim at understanding if there are moments in the circuit execution in which faults in the qubits are more critical.
For the lack of space, we only show some representative examples, being the remaining configurations very similar to the ones plotted.
%

Figure~\ref{fig_ft_tgate} shows the time dependency of QVF for the injection of a \textit{T} gate-like fault (shifts of $\theta=0$ and $\phi=\frac{\pi}{2}$) at a different time (i.e., at different circuit depths). 
Injecting this particular fault does not seem to have much variability throughout time on all the considered circuits, the QVF changes only in the range $0 - 0.2$.
We can compare this with Figure~\ref{fig_ft_ygate}, which shows the effect in time of the injection of a \textit{Y} gate-like fault ($\theta=\pi$ and $\phi=0$).
In this latter case, the impact is critical almost on all qubits at all times.
There are time instants for which the QVF for qubit 3 drops to 0, and this can be clearly deduced by the fact that injecting such a fault on qubit 3 has no effect after a certain time as that particular qubit is no longer necessary for the algorithm execution.

\begin{figure}[ht!]%
    \begin{subfigure}{.24\textwidth}
   	    \centering
        \includegraphics[width=\textwidth]{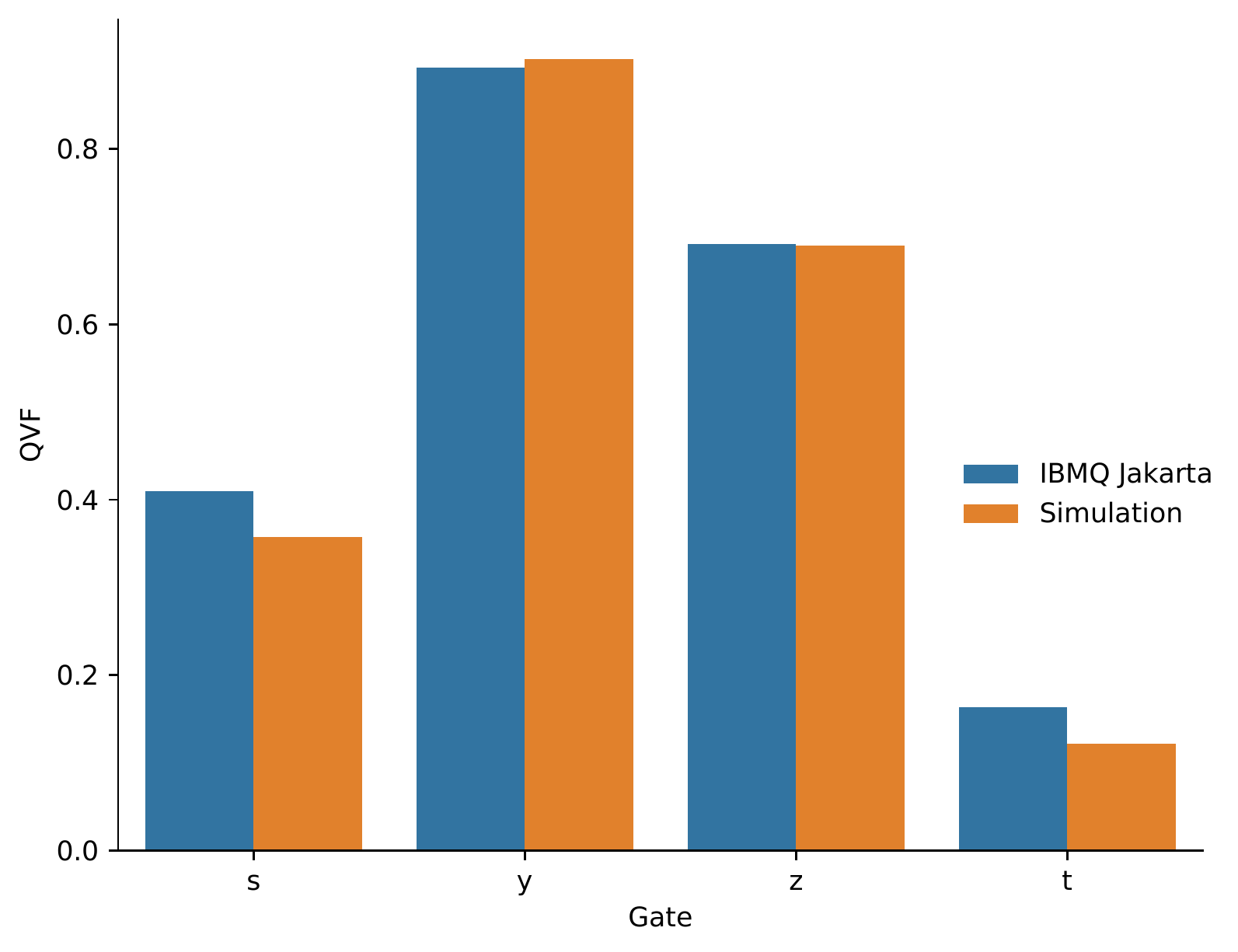}
        \caption{Bernstein-Vazirani.}
        \label{fig_hm_bv_real}
    \end{subfigure}%
    \begin{subfigure}{.24\textwidth}
        \centering
        \includegraphics[width=\textwidth]{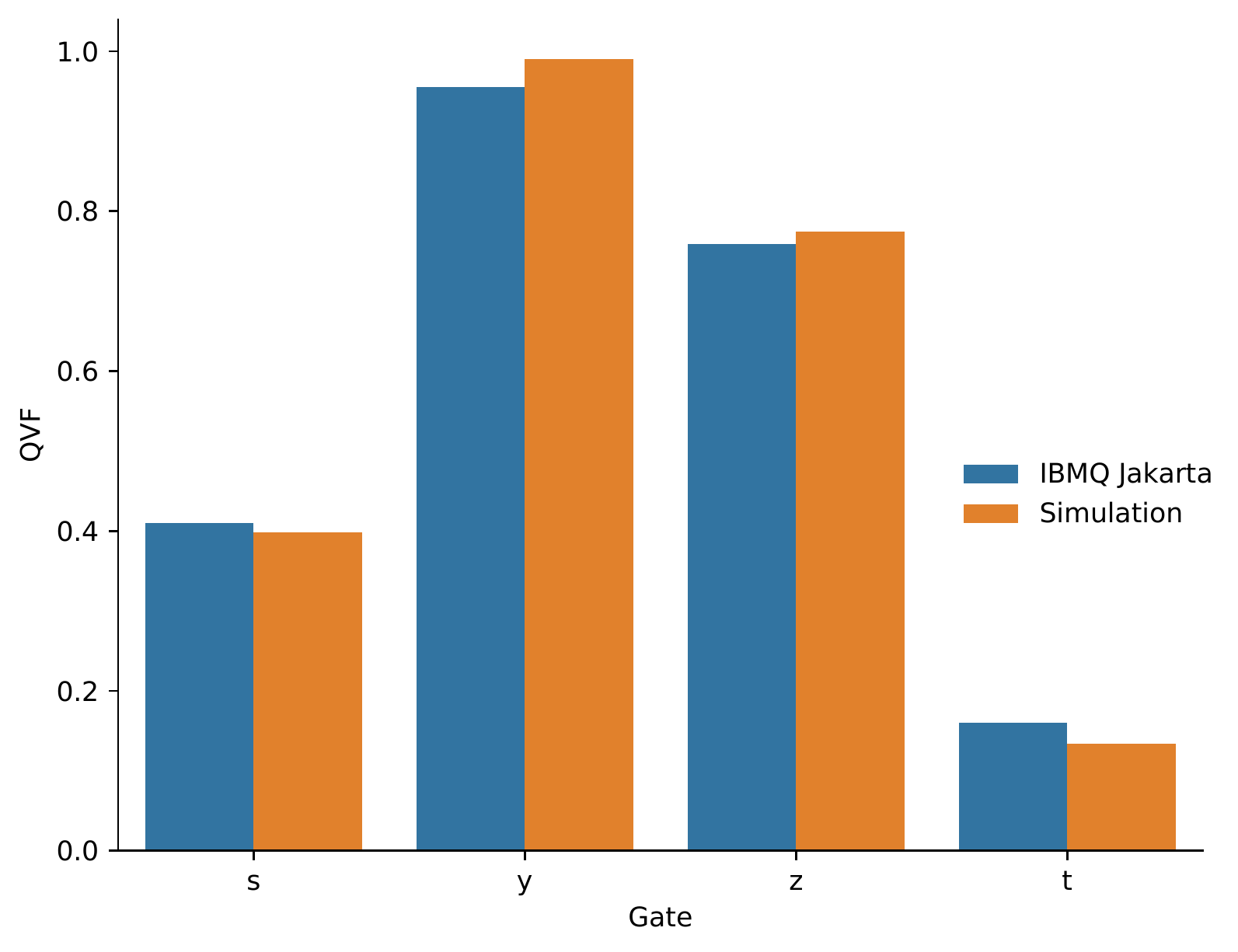}
        \caption{Grover.}
        \label{fig_hm_grover_real}
    \end{subfigure}%
    \caption{QVF comparison between simulation using IBM-Q noise model and physical machine execution (IBM-Q Jakarta).}
    \label{fig_histograms_real_simulation}
\end{figure}

Finally, Figure~\ref{fig_histograms_real_simulation} shows the comparison between a simulation including the IBM-Q noise model and the physical IBM-Q Jakarta quantum machine. Due to time constraints for IBM physical machine reservations, we compare only four specific phase shift faults, which corresponds to basic gate operations (T, S, Z, and Y), in all possible fault positions for two circuits, Bernstein-Vazirani (Fig.~\ref{fig_hm_bv_real}) and Grover (Fig.~\ref{fig_hm_grover_real}). As we can see, there is only a small variation in QVF for both circuit and fault model types, which is expected since the noise is not static and may slightly change the state probability distribution. Thus, it is safe to assume that the results from simulation with noise models are precise enough to provide insights into physical machine executions.

\section{Discussion}
\label{sec_discussion}

In this Section we discuss the impact of our result and the potential of QVF for future development. 

\subsection{Threshold theorem and QVF}
\label{sec_threshold}
The quantum threshold theorem, as an analogue to von Neumann's threshold theorem for classical computers~\cite{neumann1966theory}, states that it is possible to perform arbitrarily long quantum computations on a faulty quantum computer if the error probability per gate is below a certain threshold~\cite{Aharonov2008}.
Recently, it has been also shown that the poly-logarithmic factor present in the standard threshold theorem is actually not needed and the factor can be reduced to a constant~\cite{Fawzi2021}. It appears that there are no physical implementation limitations for quantum computer realization, and this is an exciting result.

However, the threshold theorem needs some physically reasonable assumptions about the type of noise. Therefore, although the theorem demonstrates that a fault tolerance solution exists, such a solution needs to be implemented and is obviously engineered based on the expected error rate to avoid unnecessary overheads. Not considering radiation-induced faults risks to underestimate the error rate and to guarantee sufficient reliability it will be necessary to implement additional qubit fault tolerance~\cite{muons2021}. Identifying the qubits that are more likely to require additional fault tolerance becomes then fundamental not to overestimate the redundancy or modification to apply to the circuit to make it sufficiently reliable. QVF can therefore be very useful in order to understand the impact of asynchronous faults on a circuit. This problem will become more and more important with the future increment of available qubit numbers and Michelson contrast's based QVF definition works well even when, due to a very large number of qubits, it is no more possible to fully simulate the circuit or to run it for a number of times large enough to fully understand its probability distribution function; knowing the correct state probability and the most probable among the wrong states will suffice.

\subsection{Qubit mapping}

Multi-programming, much like in classical computation, is required to improve throughput and better utilize quantum hardware, especially in NISQ-era quantum chips~\cite{li2019tackling}. However, mapping qubits into physical ones for single or multiple quantum circuits has also an impact on the overall circuit reliability~\cite{das2019case}. This impact is caused by the limited high fidelity quantum resources, cross-talk noise, and SWAP operations inserted due to the machine topology~\cite{liu2021qucloud,  murali2020software}. The qubit QVF information we provide can be used also to improve such mappings, even in the absence of faults. To improve the quality of the circuit output, for instance, the qubits that are found more resilient to injections should be allocated to resources with lower fidelity. Recently proposed qubit mapping~\cite{liu2021qucloud,  murali2020software} could further improve the circuit reliability by considering QVF as a mean to identify the qubits that are more likely, in the presence of noise, to affect the output correctness.
\section{Conclusions}
\label{sec_conclusion}

In this paper, we have proposed the Quantum Vulnerability Factor (QVF) to better evaluate the sensitivity of qubits and quantum circuits to radiation-induced faults. The interaction with particles has been demonstrated to affect the qubit state and to be a serious problem for future quantum computers. By identifying the vulnerabilities to faults of circuits and qubits, the QVF provides useful information about the reliability characteristics of a circuit and identifies the qubits that, once corrupted, are more likely to affect the output state distribution.

Using a specially crafted fault injector, built on top of Qiskit, we have evaluated the QVF of all the qubits of three quantum circuits. We have modeled the faults as phase(s) shifts of different amplitude. Our evaluation also allows to identify the kind of faults that are more critical for a circuit or qubit.

As quantum computers capability, the number of algorithms and applications, and the availability of quantum machines increases, we expect a growing interest in the radiation sensitivity of qubits. The QVF is then an effective metric to understand faults propagation and to identify the weaknesses of qubits and circuits. In the next future, quantum computers will have thousands of qubits. If we suppose to plot a QVF map for each qubit and to put all the maps one over the other like slices, we can, by means of techniques of volume rendering, explore the behavior of the whole computer, like by means of Nuclear Magnetic Resonance it is possible to explore the human body. Moreover, as our framework is also able to model correlated radiation-induced faults in multiqubit circuits, we plan to extend our evaluation to this phenomenon that, according to~\cite{muons2021}, is going to be common in quantum computers.




\bibliographystyle{IEEEtran}
\bibliography{main.bbl}

\begin{IEEEbiography}[{\includegraphics[width=1in,height=1.25in,clip,keepaspectratio]{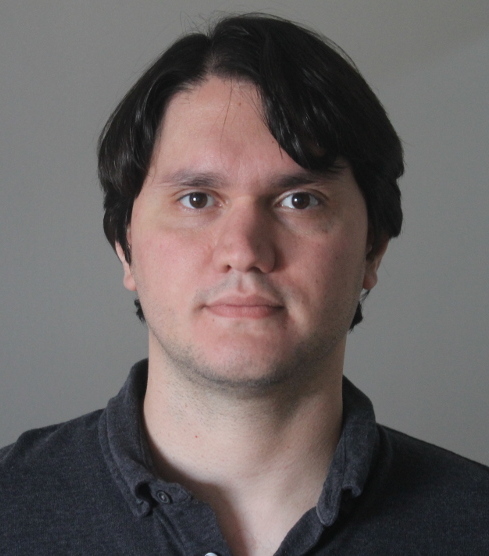}}]{Daniel Oliveira}
 received his M.S. and Ph.D. degrees from Federal University of Rio Grande do Sul (UFRGS), Porto Alegre, Brazil, in 2013 and 2018, respectively. He is currently an assistant professor at Federal University of Paran\'{a} (UFPR) in Brazil. His current research interests include fault tolerance for HPC systems and quantum computing.
\end{IEEEbiography}

\begin{IEEEbiography}[{\includegraphics[width=1in,height=1.25in,clip,keepaspectratio]{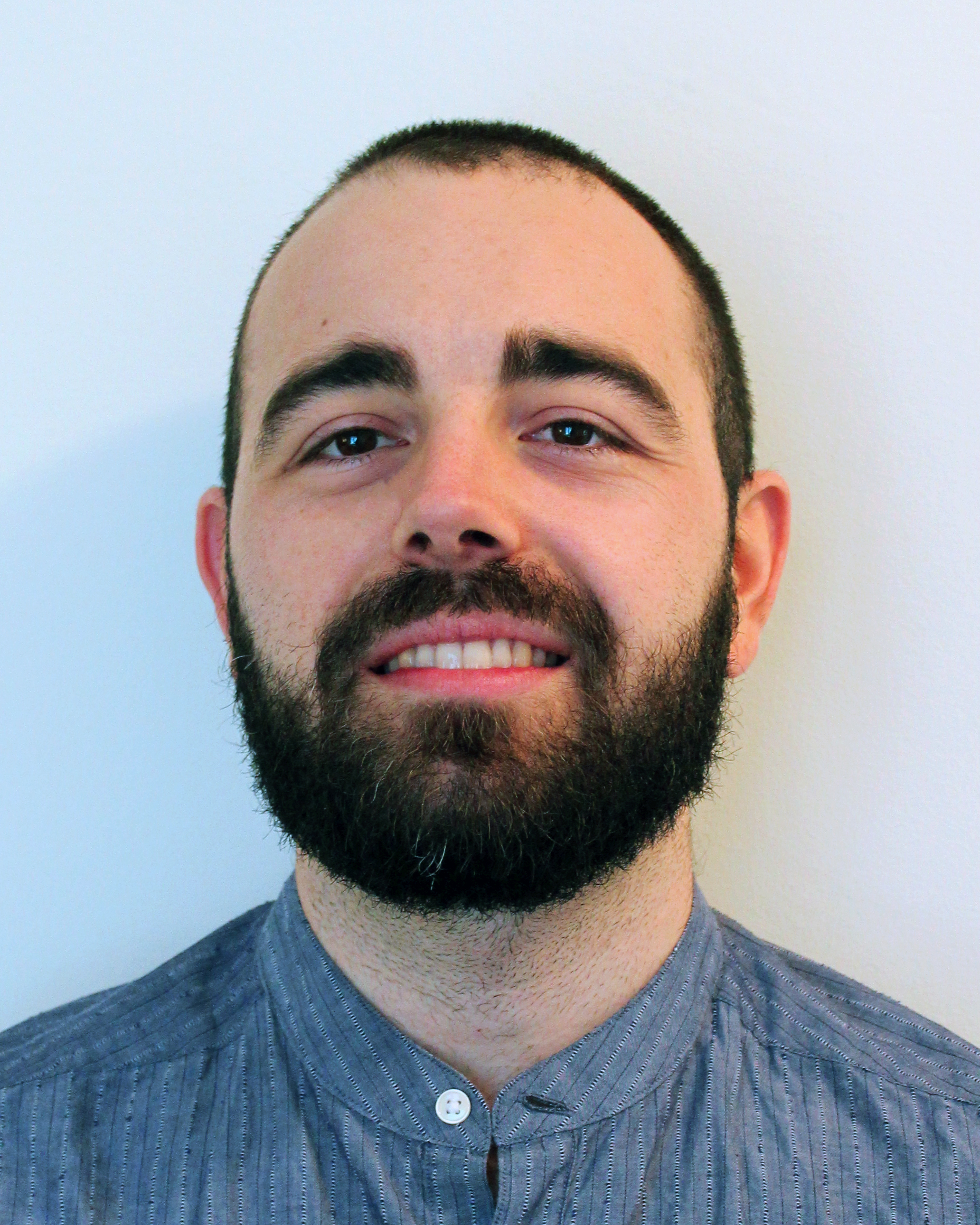}}]{Edoardo Giusto}
 obtained the M.S. degree in 2017 and Ph.D. degree in 2021 from Politecnico di Torino. He is currently a Research Assitant at the Department of Control and Computer Engineering at Politecnico di Torino. His research interests include WSNs, IoT, Smart Societies and Quantum Computing.
\end{IEEEbiography}

\begin{IEEEbiography}[{\includegraphics[width=1in,height=1.25in,clip,keepaspectratio]{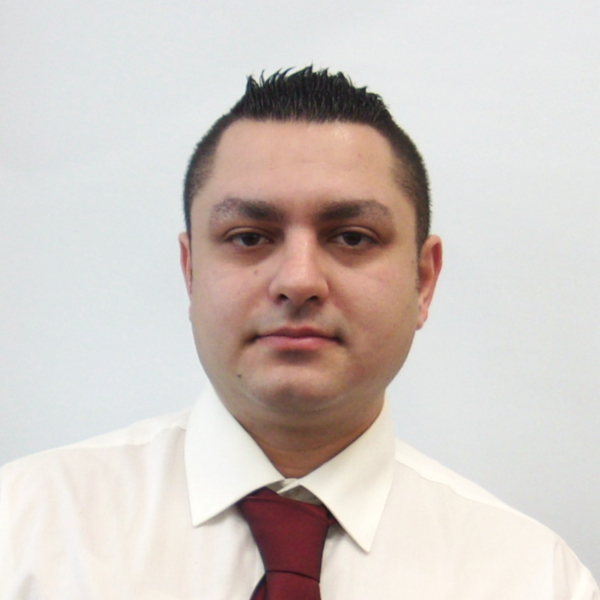}}]%
{Betis Baheri}
Betis Baheri received his B.S. degree in Computer Science from Kent State University, in 2018, and the M.S in computer science from Kent State University in 2020. His area of research while he was in Undergraduate was security and privacy. For his master he focused on HPC systems. Previously he was working on HPC scheduler and currently he is pursuing Ph.D. degree in Computer Science at same university in quantum computing and HPC systems. His main research focus is quantum error correction, quantum deep learning, and quantum machine learning on NISQ and Ion based quantum computers.  
\end{IEEEbiography}

\begin{IEEEbiography}[{\includegraphics[width=1in,height=1.25in,clip,keepaspectratio]{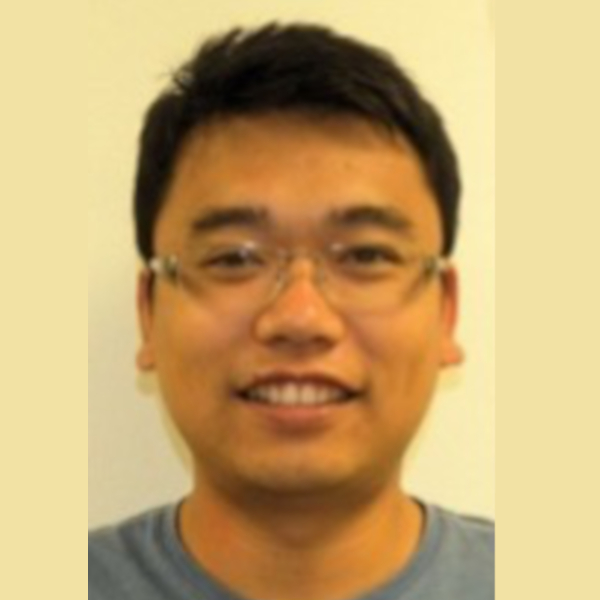}}]%
{Qiang Guan}
Dr. Qiang Guan is an assistant professor in Department of Computer Science at Kent State University, Kent, Ohio. Dr. Guan is the director of Green Ubiquitous Autonomous Networking System lab (GUANS). He is also a member of Brain Health Research Institute (BHRI) at Kent State University. He was a computer scientist in Data Science at Scale team at Los Alamos National Laboratory before joining KSU. His current research interests include fault tolerance design for HPC applications; HPC-Cloud hybrid system; virtual reality; quantum computing systems and applications. 
\end{IEEEbiography}

\begin{IEEEbiography}[{\includegraphics[width=1in,height=1.25in,clip,keepaspectratio]{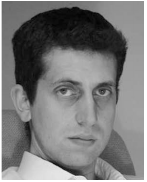}}]{Bartolomeo Montrucchio} received the M.S.
degree in electronic engineering and the Ph.D.
degree in computer engineering from the Politecnico
di Torino, Turin, Italy, in 1998, and 2002, respectively.
He is currently an Associate Professor of
Computer Engineering with the Department of Control and Computer Engineering, Politecnico di Torino. His
current research interests include image analysis and
synthesis techniques, scientific visualization, sensor
networks, RFIDs and quantum computing. 
\end{IEEEbiography}

\begin{IEEEbiography}[{\includegraphics[width=1in,height=1.25in,clip,keepaspectratio]{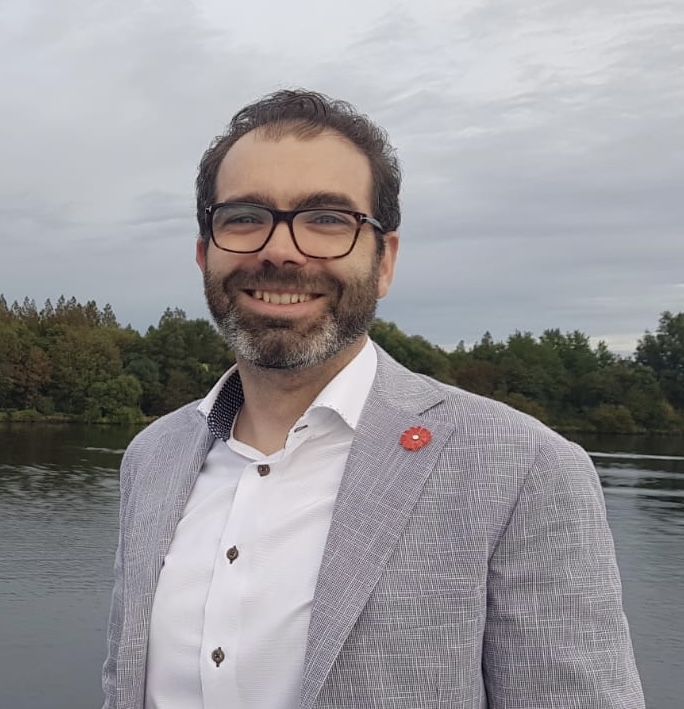}}]{Paolo Rech}
received his master and Ph.D. degrees from Padova University, Padova, Italy, in 2006 and 2009, respectively. Since 2012 Paolo is an associate professor at UFRGS in Brazil. He is the 2019 Rosen Scholar Fellow at the Los Alamos National Laboratory, he received the 2020 impact in society award from the Rutherford Appleton Laboratory, UK. Since 2020 Paolo is a Marie Curie Fellow at Politecnico di Torino. His main research interests include the evaluation and mitigation of radiation-induced effects in large-scale HPC centers and in autonomous vehicles for automotive applications and space explorations, the dependability of AI, and quantum computing.
\end{IEEEbiography}

\end{document}